\documentclass[fleqn,usenatbib]{mnras}
\usepackage{lscape}
\usepackage{rotating}
\usepackage{amssymb}
\usepackage{float}
\usepackage{hyperref}
\usepackage{newtxtext,newtxmath}
\usepackage{multirow}

\usepackage{xcolor}

\usepackage{etoolbox}
\makeatletter
\patchcmd\@combinedblfloats{\box\@outputbox}{\unvbox\@outputbox}{}{%
    \errmessage{\noexpand\@combinedblfloats could not be patched}%
}%
\makeatother

\def\sm{\hbox{S$_{\rm M}$}}

\def\msun{\hbox{M$_\odot$}}

\def\mstar{\hbox{$M_\star$}}
\def\mcstar{\hbox{$M_{\rm c, \star}$}}

\def\mgc{\hbox{$M_{\rm GC}$}}
\def\ngc{\hbox{$N_{\rm GC}$}}

\def\mhalo{\hbox{$M_{\rm halo}$}}
\def\m200{\hbox{$M_{\rm 200}$}}

\def\t4{\hbox{$t_{\rm 4}$}}

\def\cm3{\hbox{cm$^{-3}$}}

\input psfig.sty
\input epsf.sty
\usepackage{graphicx}
\usepackage{epsfig}
\title[\mgc-\mhalo\ in E-MOSAICS]
{The globular cluster system mass-halo mass relation in the E-MOSAICS simulations}
\author[Bastian et al.] {Nate Bastian,$^{1}$, Joel Pfeffer,$^{1}$ J.~M.~Diederik Kruijssen,$^{2}$ Robert A.~Crain,$^{1}$   \newauthor  Sebastian Trujillo-Gomez$^{2}$, and Marta Reina-Campos$^{2}$\\ 
$^{1}$Astrophysics Research Institute, Liverpool John Moores University, 146 Brownlow Hill, Liverpool L3 5RF, UK\\
$^{2}$Astronomisches Rechen-Institut, Zentrum f\"ur Astronomie der Universit\"{a}t Heidelberg, M\"onchhofstrasse 12-14, D-69120 Heidelberg, Germany.\\
}

\date{Accepted. Received; in original form}
\pagerange{\pageref{firstpage}--\pageref{lastpage}}
\pubyear{2020}
\begin{document}
\maketitle
\label{firstpage}
\begin{abstract}
Linking globular clusters (GCs) to the assembly of their host galaxies is an overarching goal in GC studies. The inference of tight scaling relations between GC system properties and the mass of both the stellar and dark halo components of their host galaxies are indicative of an intimate physical connection, yet have also raised fundamental questions about how and when GCs form. Specifically, the inferred correlation between the mass of a GC system (\mgc) and the dark matter halo mass (\mhalo) of a galaxy has been posited as a consequence of a causal relation between the formation of dark matter mini-haloes and GC formation during the early epochs of galaxy assembly. We present the first results from a new simulation of a cosmological volume ($L=34.4$~cMpc on a side) from the E-MOSAICS suite, which includes treatments of the formation and evolution of GCs within the framework of a detailed galaxy formation model. The simulated \mgc-\mhalo\ relation is linear for halo masses $>5\times10^{11}~\msun$, and is driven by the hierarchical assembly of galaxies. Below this halo mass, the simulated relation features a downturn, which we show is consistent with observations, and is driven by the underlying stellar mass-halo mass relation of galaxies. 
Our fiducial model reproduces the observed \mgc-\mstar\ relation across the full mass range, which we argue is more physically relevant than the \mgc-\mhalo\ relation. We also explore the physical processes driving the observed constant value of $\mgc / \mhalo \sim 5\times10^{-5}$ and find that it is the result of a combination of cluster formation physics and cluster disruption.

\end{abstract}

\section{Introduction}
\label{sec:intro}

One of the main goals in stellar cluster research is to place globular clusters (GCs) and their young counterparts, young massive clusters (YMCs), in the wider context of galaxy assembly (see e.g., the recent reviews of \citealt{Kruijssen_14} and \citealt{forbes_review_18}). How do GCs trace the stellar component and dark halo mass of their host galaxy, and what drives these relations?  A number of works have highlighted that GC populations obey scaling relations with the properties of their host galaxies, suggesting an intimate and causal relationship between the two \citep[e.g.,][]{Peng_et_al_08, Georgiev_et_al_10}.  Particular attention has been placed on the relation between the mass of a GC population (\mgc) and that of the host galaxy's halo or virial mass (\mhalo\ or \m200, which we use interchangeably). Observations show a near-linear scaling between the mass in these two quantities \citep[e.g.,][]{blakeslee97, blakeslee99} with a proportionality factor $\eta = \mgc / \mhalo \sim 5\times10^{-5}$ \citep[e.g.,][]{Harris_et_al_17}.  
At first glance, such a relation appears surprising, given that the more natural connection is expected to be that between the GCs and the stellar mass ($\mstar$) of the host, if GC formation is a natural consequence of star formation.  Hence, the relation $\sm$ ($=100\times \mgc / \mstar$) versus \mstar\ would be expected to encode more information about GC formation and their relation to the host galaxy.  Additionally, there exists a non-linear relation between the \mstar\ and the inferred \mhalo\ of galaxies \citep[e.g.][]{Behroozi_et_al_13,Moster_et_al_13}, rendering the apparently simple relation between \mgc\ and \mhalo\ relation all the more surprising.

Some authors have taken the empirical \mgc-\mhalo\ relation to conclude that GCs must have formed early on in the galaxy assembly process, before baryonic processes (e.g., feedback) cause the star formation rate (SFR) to deviate strongly from the gas inflow rate \citep[see e.g.][]{van_de_voort_11}.  The implication would be that GCs are more directly tied to the dark matter of the host galaxy than the stars, forming in dark matter mini-haloes during the early epoch of galaxy assembly. For example, if a given mass of dark matter halo results in a fixed amount of mass/number of GCs to form, then the linear \mgc-\mhalo\ relation would be a natural result \citep[e.g.,][]{Boylan-Kolchin_17}. Within cosmological simulations, the number of independent dark-matter mini-haloes that are capable of forming GCs but have not yet formed stars decreases strongly with decreasing redshift, implying that if mini-haloes were the preferential site of GC formation, most GCs must form before $z=6$.  However, there is growing evidence that GC formation is not restricted to the early Universe, and that they can form across all cosmic history, from very early times, to cosmic noon and even in the local Universe today \citep[e.g.,][]{Holtzman_et_al_92, Schweizer_and_Seitzer_98, Kruijssen_15, Vanzella_et_al_17, Johnson_et_al_17, Reina-Campos_et_al_19, Usher_et_al_19}.

An alternative explanation of the linear \mgc-\mhalo\ relation was put forward by \citet{Kruijssen_15}, who argued that the combination of the physics of cluster formation and the preferential disruption of GCs based on their environment would result in a near-linear relation.  The subsequent merging of galaxies (along with their GC populations) would then act to fully linearise the relation.  A related interpretation has recently been put forward by \citet{El-Badry_et_al_19} and \citet{Choksi_and_Gnedin_19}.  These authors suggest that the hierarchical growth of galaxies naturally leads to such a tight correlation, independent of the adopted cluster formation model, except at the low halo mass end, which may retain some memory of GC formation.  Effectively, in this view, the \mgc-\mhalo\ relation is a result of the central limit theorem, echoing the interpretation of the origin of scaling relations between central supermassive black holes and galaxy properties advanced by \citet{jahnke_and_maccio_11}. In both the \citet{El-Badry_et_al_19} and \citet{Choksi_and_Gnedin_19} works the authors show that a GC formation model where clusters form throughout cosmic history can result in the observed \mgc-\mhalo\ relation, without any explicit connection between the formation of GCs and dark matter (mini) haloes.

While these works were able to successfully explain the linear relation between the mass in globular clusters and that of the host halo, due to the lack of spatial resolution and baryonic physics in the models the authors were unable to explore the normalisation of the relation.  What physical processes are responsible for GCs to represent a near constant mass fraction of 0.005\% of their host galaxy?  Is cluster formation or disruption the dominant process?  If it is the former, what is the relative role of the shape of the cluster mass function and the cluster formation efficiency?

Observational studies have attempted to trace the \mgc-\mhalo\ relation to ever lower halo masses in order to test if, and at what mass scale, the linear relation breaks down.  \citet{Forbes_et_al_18} have studied a heterogeneous sample of nearby dwarf galaxies to extend the relation down to $\mhalo \sim 10^{8}$~\msun, and conclude that the relation continues to be linear down to at least this limit.  This is in tension with the models of \citet{El-Badry_et_al_19} and \citet{Choksi_and_Gnedin_19}, which predict a downturn near $\sim5\times10^{11}$~\msun.  The origin of this discrepancy is not entirely clear -- it is one of the goals of this paper to understand this difference.  We note, however, that if galaxies without GCs are included in the \citet{Forbes_et_al_18} sample, the running median of the \mgc-\mhalo\ relation does show a downturn near $\mhalo \sim 10^{10}$~\msun\ \citep[see][]{Georgiev_et_al_10}.

In the present work, we study the \mgc-\mhalo\ relation in the E-MOSAICS simulations of the co-formation and evolution of GCs and their host galaxies in a fully cosmological framework (\citealt{Pfeffer_et_al_18}; \citealt{Kruijssen_et_al_19a}; Crain et al.\ in prep.).  With these simulations, we can trace the buildup of full GC populations alongside their host galaxy and explore the role of various physical processes in setting their properties.  Additionally, we can analyse the simulations directly (i.e., measuring the dark matter halo mass of each of the galaxies in the simulations) or consider observational proxies commonly used in the literature (i.e., measure the stellar mass of a galaxy and translate this to a halo mass using a scaling relation). This allows us to explore many of the underlying assumptions in observational studies of the \mgc-\mhalo\ relation, as well as to investigate how it relates to other observed correlations between GC and galactic properties.

This paper is organised as follows.  In Section~\ref{sec:emosaics} we introduce the simulations used throughout this work.  In Section~\ref{sec:results} we present the main results and interpretation from the simulations, namely the origin of the \mgc-\mhalo\ relation, comparisons with observations and other scaling relations like the \mgc-\mstar\ relation.  Finally, in Section~\ref{sec:discussion} we discuss the results and present our conclusions.

\section{The E-MOSAICS simulations}
\label{sec:emosaics}

\subsection{Simulation setup}

The E-MOSAICS (MOdelling Star cluster population Assembly In Cosmological Simulations within EAGLE) project is a suite of cosmological hydrodynamical simulations based on the EAGLE (Evolution and Assembly of GaLaxies and their Environments) galaxy formation model \citep{Schaye_et_al_15, Crain_et_al_15} with a subgrid treatment of stellar cluster formation, evolution and disruption \citep{Kruijssen_et_al_11, Pfeffer_et_al_18}.  The physical ingredients of the subgrid stellar cluster models have been presented in detail in \citet{Pfeffer_et_al_18} and \citet{Kruijssen_et_al_19a} and we refer the interested reader to those papers for more details.  In the present work we use a new suite of simulations, with the same physical model for cluster formation and evolution and at the same EAGLE resolution as in previous works, but instead of focusing on zoom-in simulations of Milky Way-mass haloes \citep{Pfeffer_et_al_18, Kruijssen_et_al_19a} or a small periodic volume \citep[$L=12.5$~cMpc][]{Pfeffer_et_al_19b}, we use the simulation of a large periodic volume of $34.4$~comoving Mpc (cMpc) on a side.  This simulation has a volume $2.6$ times larger than the previous largest EAGLE simulation at the same resolution (L025N0752) and will be presented in detail in Crain et al.~(in prep.).

EAGLE is a suite of hydrodynamical simulations of galaxy formation in the $\Lambda$ cold dark matter cosmogony \citep[for full details, see][]{Schaye_et_al_15, Crain_et_al_15}.
The simulations are evolved with a highly modified version of the $N$-body, smoothed particle hydrodynamics code \textsc{Gadget3} \citep[last described by][]{Springel_05}, which include subgrid routines describing radiative cooling \citep{Wiersma_Schaye_and_Smith_09}, star formation \citep{Schaye_and_Dalla_Vecchia_08}, stellar evolution and mass-loss \citep{Wiersma_et_al_09}, the seeding and growth of black holes (BHs) via gas accretion and BH-BH mergers \citep{Rosas-Guevara_et_al_15}, and feedback associated with star formation and BH growth \citep{Booth_and_Schaye_09}.
The parameters describing the energy feedback from supernovae and active galactic nuclei are calibrated such that the simulations reproduce the present-day galaxy stellar mass function, size-mass relation of disc galaxies and the relation between the mass of central BHs and galaxy stellar mass.
The simulations were performed assuming a \citet{Planck_2014_paperXVI} cosmology, with $\Omega_\mathrm{m} = 0.307$, $\Omega_\mathrm{\Lambda} = 0.693$, $\Omega_\mathrm{b} = 0.04825$, $h = 0.6777$ and $\sigma_8 = 0.8288$.

Coupled to the EAGLE model is the MOSAICS model describing the formation and evolution of star clusters \citep{Kruijssen_et_al_11, Pfeffer_et_al_18}.
Star clusters are treated as a subgrid component of the stellar particles, such that they
adopt the properties of the host particle (i.e., positions, velocities, ages, abundances) and form and evolve according to local properties within the simulation (namely the local ambient gas and dynamical properties).
Cluster formation within the model is described by two main parameters, the cluster formation efficiency \citep[CFE, the fraction of stars formed within bound clusters,][]{Bastian_08} and the upper exponential truncation to the \citet{Schechter_76} cluster mass function ($M_{\rm c, \star}$, with a power-law index of $-2$ at lower masses).
The fiducial E-MOSAICS cluster formation model allows the CFE and \mcstar to vary as a function of the local environmental conditions;  specifically we adopt the \citet{Kruijssen_12} model for the CFE and the \citet{Reina-Campos_and_Kruijssen_17} model for $M_{\rm c, \star}$.
Alternative cluster formation models were tested by adopting a constant CFE or a pure power-law mass function (discussed in further detail below).
In order to reduce memory requirements for the simulations, only clusters with initial masses $>5\times 10^3$~\msun\ are evolved and we assume instant disruption for clusters formed with lower masses.
Following their formation, clusters may lose mass through stellar evolution (according to the EAGLE model), two-body relaxation depending on the strength of the local tidal field\footnote{Following \citet{Gieles_and_Baumgardt_08}, we have added a term to the mass-loss rate from two-body relaxation to account for `isolated' clusters, i.e. those in a weak tidal field. This change has only a minor influence, mainly for clusters with masses $\lesssim10^4$~\msun.} and tidal shocks from rapidly changing tidal fields based on the derivations of \citet{Gnedin_Hernquist_and_Ostriker_99}, \citet{Prieto_and_Gnedin_08}, and \citet{Kruijssen_et_al_11}.
Star clusters that fall below a mass of $10^2$~\msun\ (through any mechanism) are assumed to be fully disrupted.
Total removal of clusters via dynamical friction (assuming they merge to the centre of their host galaxy) is treated in post-processing and applied at every snapshot in the simulation \citep[see][]{Pfeffer_et_al_18}.

The E-MOSAICS project aims to carry out self-consistent simulations of the co-formation and evolution of galaxies along with their stellar cluster populations.  Specifically, we are aiming to test whether the young massive clusters observed in nearby galaxies (as well as our own) share the same formation mechanisms as the ancient GCs.  Using E-MOSAICS, we have carried out a number of studies which have attacked this problem from a variety of angles.  For example, the simulations have been shown to reproduce the observed scaling relations for young massive clusters in nearby galaxies, including systematic variations in the cluster initial mass function and the fraction of stars that form in clusters as a function of environment \citep{Pfeffer_et_al_19b}. They have been used to investigate the origin of the ``blue tilt" in GC populations without invoking multiple epochs of star formation \citep{Usher_et_al_18}, as well as the galaxy-to-galaxy scatter in the age-metallicity relation of GCs \citep{Kruijssen_et_al_19a}.  The latter  resulted in the inference of a previously unknown major accretion event during the early assembly of the Milky Way \citep{Kruijssen_et_al_19b,Kruijssen_et_al_20}.  \citet{Hughes_et_al_19} used the simulations to trace the build up of galaxy halos (and their GC populations) using stellar streams of accreted satellites.  They predict that GCs belonging to identifiable stellar streams should be, on average, younger than GCs located off streams, which has recently been observed in the M31 GC system \citep{Mackey_et_al_19}.  \citet{Reina-Campos_et_al_19} used the simulations to show that, at least for Milky Way-like galaxies, the GC populations are expected to form across a wide range of redshifts, with a peak GC formation rate at $z\sim2$. Finally, we have used the E-MOSAICS to quantify the amount of dynamical mass loss experienced by GCs \citep{Reina-Campos_et_al_18}, allowing us to reproduce the fractions of the stellar bulge \citep{Hughes_et_al_20} and stellar halo that are constituted by disrupted GCs \citep{Reina-Campos_et_al_20}. Given that the simulations are able to reproduce a broad range of observational properties (while also making explicit predictions for future observables) of both young stellar clusters and old GCs, we argue that the basic model adopted by E-MOSAICS (i.e.\ that the same underlying physical mechanisms govern cluster formation across cosmic history) is accurate and can be further applied to new regimes.

\subsection{E-MOSAICS periodic volume}
\label{sec:volume}

We conducted a simulation (L034N1034) of a periodic cube of size $L = 34.4$~cMpc on a side with the E-MOSAICS model.
Full details of the simulation will be presented in Crain et al.~(in prep.).
Briefly, the simulation adopts the same sub-grid physics as for the E-MOSAICS zoom-in simulations \citep{Pfeffer_et_al_18, Kruijssen_et_al_19a}.
The simulation uses $2 \times 1034^3$ particles (with an equal number of baryonic and dark matter particles), such that the dark matter particle mass is $1.21\times 10^6$~\msun\ and the initial gas particle mass is $2.26 \times 10^5$~\msun. 
The Plummer-equivalent gravitational softening length is fixed in comoving units to 1/25 of the mean interparticle separation (1.33 comoving kpc) until z = 2.8, and in proper units (0.35 pkpc) thereafter.
The simulation was performed with the `Recalibrated' EAGLE model \citep[see][]{Schaye_et_al_15}, using a resolution identical to the L025N0752 volume, and thus achieves results for the galaxy population consistent with the EAGLE Recal-L025N0752 simulation.
In total, 29 snapshots and 405 `snipshots' were saved between redshifts $z=20$ and $z=0$ (spaced approximately linearly in scale factor).

Unlike the previous E-MOSAICS simulations, which re-ran simulations with different cluster formation physics \citep[see][]{Pfeffer_et_al_18}, for the L034N1034 simulation we ran all four cluster formation models (Section \ref{sec:formation_models}) in parallel.
This is possible since the EAGLE galaxy formation model is independent of the MOSAICS star cluster model.
Once formed, all star clusters are then evolved following the same cluster evolution model according to their local tidal field (Section \ref{sec:emosaics}).
We note that it is necessary to run the MOSAICS model on the fly, because the variation of the local tidal field at the position of each cluster must be followed at a time resolution of $<1$~Myr, which is much finer than the computationally feasible output interval of simulation snapshots.

Galaxies (subhaloes) were identified in the simulation using the method described in \citet{Schaye_et_al_15}.
Dark matter structures were first identified using the friends-of-friends (FoF) algorithm \citep{Davis_et_al_85} with a linking length $0.2$ times the mean interparticle separation.
Gravitationally bound substructures (galaxies and subhaloes) were then identified using the \textsc{subfind} algorithm \citep{Springel_et_al_01, Dolag_et_al_09}.
Within each FoF group, the galaxy that contains the particle with the lowest value of the gravitational potential is considered to be the \textit{central} galaxy, and all other galaxies are considered to be \textit{satellite} galaxies.
Stellar particles (and the star clusters they host) are thus allocated to galaxies by the \textsc{subfind} binding criteria.
At $z=0$, the volume has $[465, 69, 7]$ haloes with $M_{200} > [10^{11}, 10^{12}, 10^{13}]$~\msun.

\subsection{Alternate cluster formation models}
\label{sec:formation_models}

\defcitealias{Reina-Campos_et_al_19}{RC19}

In order to explore the effect of GC formation physics on the resulting \mgc-\mhalo\ relation, we follow \citet[hereafter \citetalias{Reina-Campos_et_al_19}]{Reina-Campos_et_al_19} who investigated the four different E-MOSAICS cluster formation models, which are outlined below as well as in Table~\ref{tab:models}.

\begin{itemize}
\item {\bf Fiducial Model:}  This is our default model of which the CFE ($\Gamma$) and the truncation mass of the Schechter initial cluster mass function (ICMF) vary as a function of the properties of the local environment where the stars/clusters are forming.
\item {\bf CFE only:} This was referred to as the ``$\alpha=-2$" model in \citetalias{Reina-Campos_et_al_19}.  In this model, the ICMF is invariant as a pure power-law function with an index of $-2$, although it is sampled stochastically.  The CFE, like in the fiducial model, varies with the local environment.
\item {\bf M$_{\rm c, \star}$ only:} This was referred to as the ``$\Gamma = 10$\%" model in \citetalias{Reina-Campos_et_al_19}.  This model sets the CFE at a constant level (10\%) and allows the truncation mass ($\mcstar$) of the ICMF to vary as in the fiducial model.
\item {\bf No formation physics:} In this model both the CFE and the ICMF are invariant (fixed at 10\% and a pure power-law with an index of $-2$, respectively).
\end{itemize}

\begin{table}
\caption{Cluster formation models considered in this work. From left to right, columns contain the name of the cluster formation scenario and the description used for the CFE and the ICMF, respectively.} 
\label{tab:models}      
\centering  			
\resizebox{\hsize}{!}{   
\begin{tabular}{l|c|c}
 Name & CFE & ICMF \\ \hline
\multirow{3}{*}{Fiducial} & $\Gamma(\Sigma, Q, \kappa)$ & Schechter function ($\alpha=-2$) and  \\
& \citet{Kruijssen_12} &  $M_{\rm cl,max}(\Sigma, Q, \kappa)$ \\
   & & \citet{Reina-Campos_and_Kruijssen_17} \\ \hline

\multirow{ 2}{*}{CFE only} & $\Gamma(\Sigma, Q, \kappa)$ &  Power-law of index \\
& \citet{Kruijssen_12} & $\alpha=-2$ \\\hline

\multirow{3}{*}{\mcstar\ only}   & \multirow{3}{*}{$\Gamma = 10\%$} & Schechter function ($\alpha=-2)$ and \\
    & & $M_{\rm cl,max}(\Sigma, Q, \kappa)$ \\
  & & \citet{Reina-Campos_and_Kruijssen_17} \\\hline

\multirow{ 2}{*}{No formation physics} & \multirow{ 2}{*}{$\Gamma = 10\%$} & Power-law of index \\
& & $\alpha=-2$ \\ \hline \hline

\end{tabular}}
\end{table}

\subsection{Analysis}
\label{sec:analysis}

We focus most of our analysis on central galaxies, because satellite galaxies can have their dark matter haloes (and GC populations) significantly affected by interactions with the central host.  Most observational studies use scaling relations to infer the dark matter halo mass from the observed stellar mass, which are calibrated primarily on central galaxies, implying that the inclusion of satellites from our models would not be consistent.
We limit our analysis to well-resolved galaxies with stellar masses $>10^8$~\msun\ ($\gtrsim 500$ stellar particles), similar to the galaxy mass limit in \citet{Peng_et_al_08}.
At $z=0$, this gives us a sample of 992 central galaxies, and 1707 galaxies in total when including satellites.

We refer to Appendix~\ref{sec:gc_selection} for details on how the GC sample was selected in the present work, as well as on how this affects the results.  Briefly, we determine the total mass of the GC population using the top two decades of the present day GC mass function for each galaxy.

\section{Results}
\label{sec:results}

\begin{figure*}
\centering
\includegraphics[width=14.25cm]{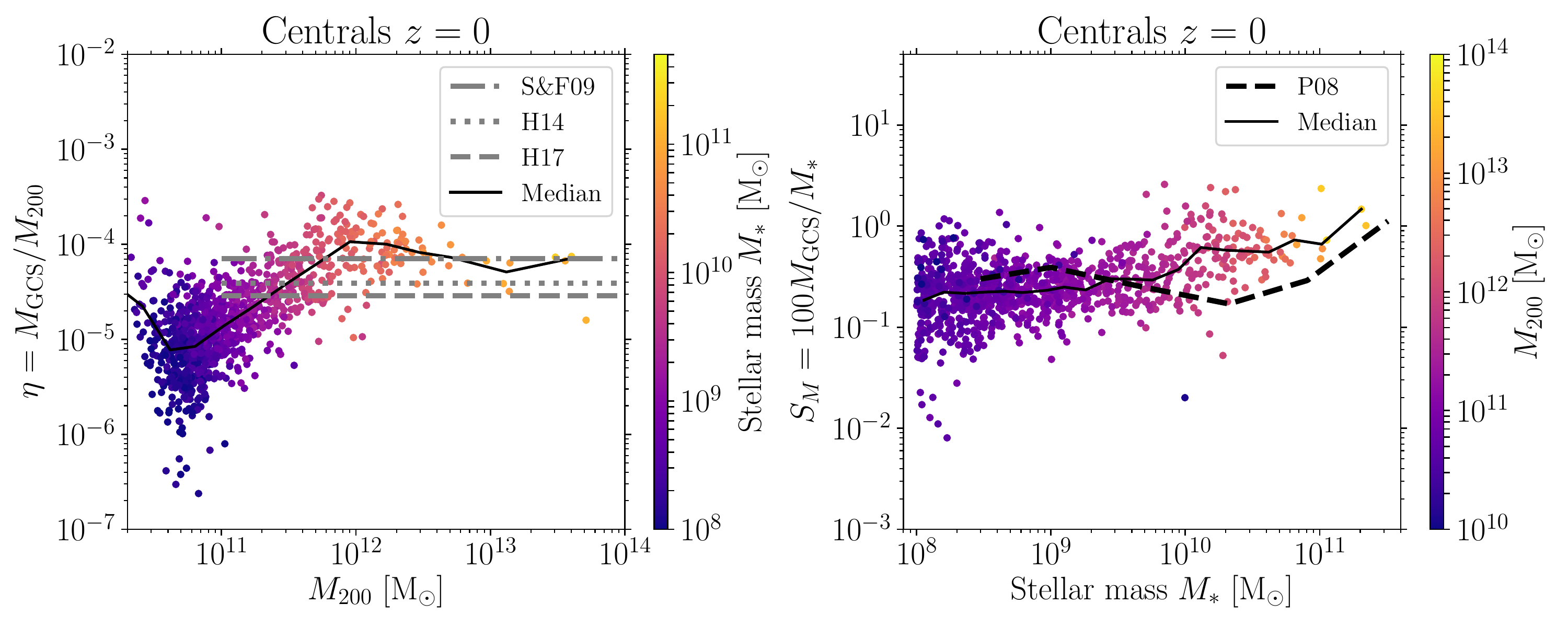}
\includegraphics[width=14.25cm]{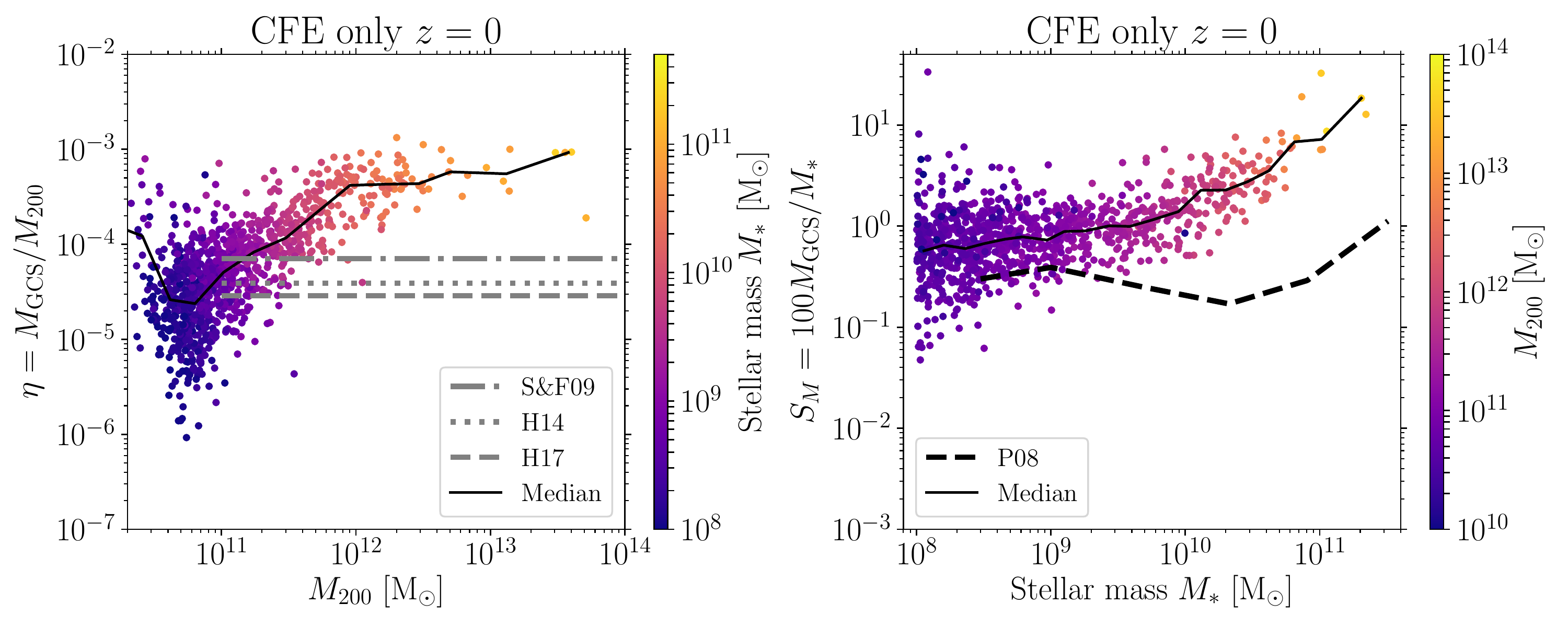}
\includegraphics[width=14.25cm]{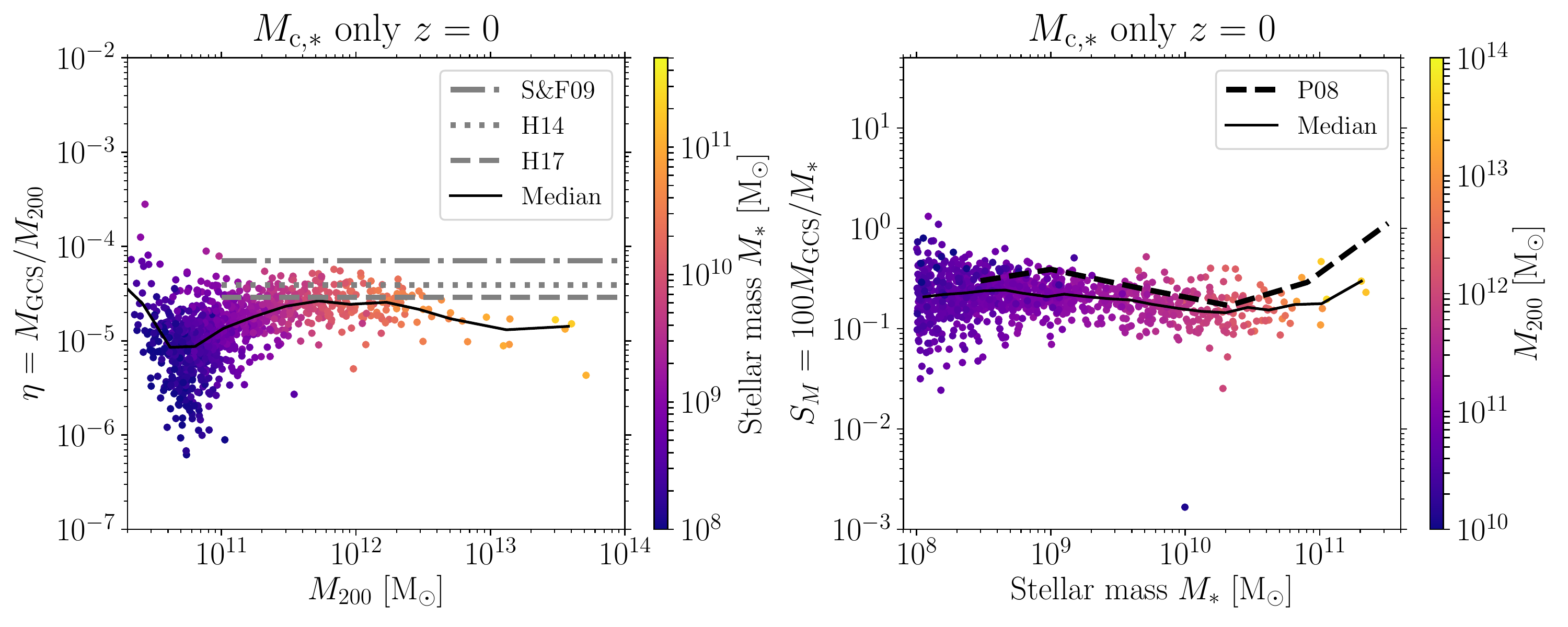}
\includegraphics[width=14.25cm]{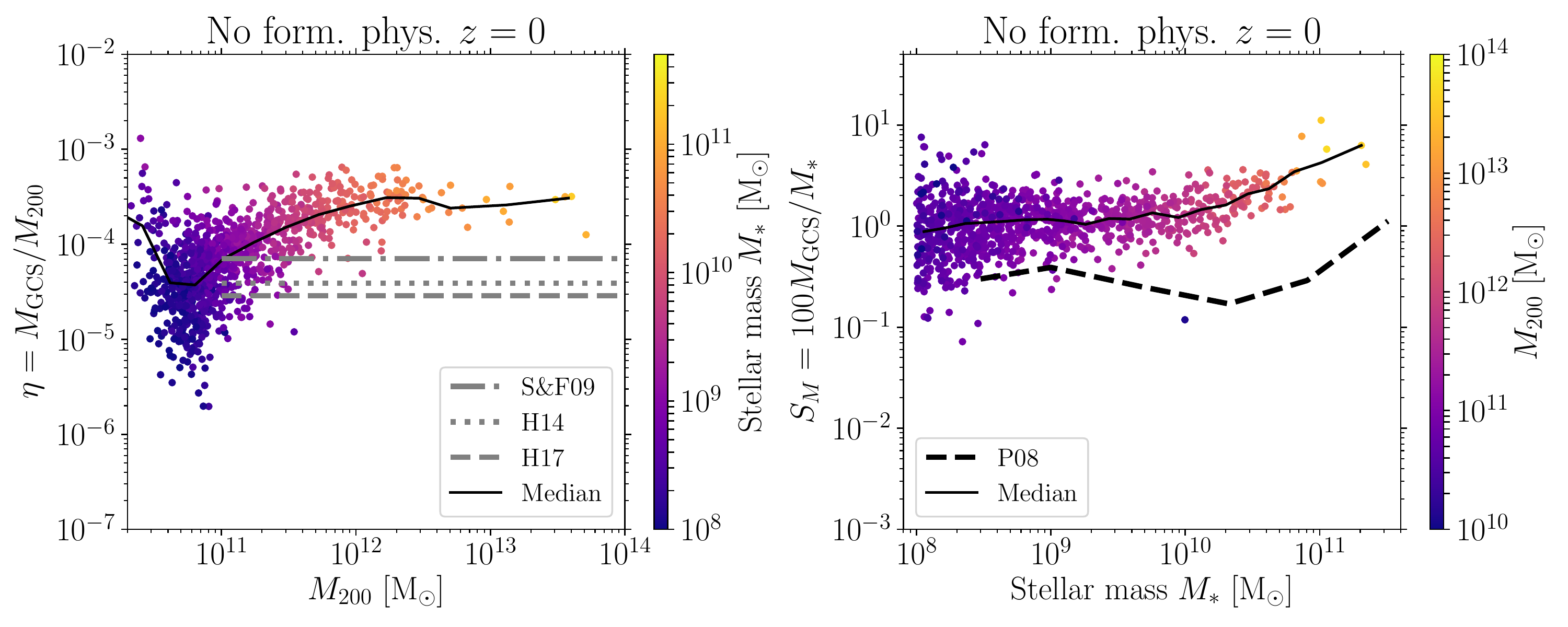}

\caption{The \mgc-\mhalo\  (left panels) and \sm-\mstar\ (right panels) relations for galaxies within our simulation. See Table~\ref{tab:models} for a summary of the differences between the simulations. The solid lines show the running median while the dashed, dotted and dash-dotted lines show observational results from the literature (see text for details), which are restricted to halo masses $\gtrsim10^{11}$~\msun.}
\label{fig:eta}
\end{figure*}

\subsection{The \mgc-\mhalo\ relation}
\subsubsection{The shape of the relation}

\defcitealias{Spitler_and_Forbes_09}{S\&F09}
\defcitealias{Hudson_et_al_14}{H14}
\defcitealias{Harris_et_al_17}{H17}

The \mgc-\mhalo\ relation from the E-MOSAICS volume is shown in Fig.~\ref{fig:eta} for the fiducial model and for the three alternative cluster formation physics models in the left-hand panels.  Additionally, horizontal lines show the estimated value of $\eta$ from three observational studies, namely \citet[S\&F09; dash dotted line]{Spitler_and_Forbes_09}, \citet[H14; dotted line]{Hudson_et_al_14} and \citet[H17; dashed line]{Harris_et_al_17}. While the E-MOSAICS simulations do not invoke any direct relation between the dark matter halo of the galaxy and the number (or mass) of GCs within it, we see that for all models the \mgc-\mhalo\ relation is linear above a halo mass of $\sim5 \times 10^{11}$~\msun.  This is consistent with the models of \citet{El-Badry_et_al_19} and \citet{Choksi_and_Gnedin_19} who found that this linear behaviour is driven by the hierarchical build up of galaxies (essentially the central limit theorem) and is not tied to the formation of GCs nor their connection to dark matter.  As we find the same behaviour in all of our models, regardless of the cluster formation physics included, our results are consistent with the conclusions of these authors, namely that the observed \mgc-\mhalo\ relation does not imply a causal relation between dark matter and GCs.

However, like \citet{El-Badry_et_al_19} and \citet{Choksi_and_Gnedin_19}, we also find a downturn in the relation at lower halo masses, with the exact location sensitive to the adopted physics.  In addition, we see that the normalisation of the relation is also a strong function of the adopted physics.
These aspects are discussed in more detail in the following sections.

\subsubsection{The normalisation of the relation}
\label{sec:normalisation}
While all of our simulations find a close to linear \mgc-\mhalo\ relation above a certain halo mass, regardless of the adopted formation physics, we see that the normalisation of the relation is dependent on the adopted model.  In the rest of the discussion, we will only focus on the linear part of the relation in our simulations.   For our fiducial model, we find a median $\eta$ value that is in good agreement with \citetalias{Spitler_and_Forbes_09} and slightly higher than \citetalias{Hudson_et_al_14} and \citetalias{Harris_et_al_17}.  This slight over abundance of GCs may be caused by the under-disruption of metal rich GCs within the E-MOSAICS simulations and we point the interested reader to \citet[Appendix~D]{Kruijssen_et_al_19a} for a detailed discussion of this point.  

The simulations using other formation physics models are systematically offset from the observed value of $\eta$.  If \mcstar\ is fixed (i.e., a pure power-law cluster initial mass function with no truncation) and the CFE is allowed to vary, we end up producing too many GCs.  If instead we adopt a constant CFE (10\%) and allow \mcstar to vary, we underestimate $\eta$, meaning that we do not produce enough GCs. The former effect is dominant, because using a constant CFE in combination with a power law ICMF leads to the overprediction of $\eta$. Hence, while the hierarchical buildup of galaxies naturally leads to a linear \mgc-\mhalo\ relation for $\sim L^\star$ galaxies and above, independently of the input GC formation physics, the normalisation of the relation does encode important information on the physics of GC formation. 

\begin{figure}
\centering
\includegraphics[width=8cm]{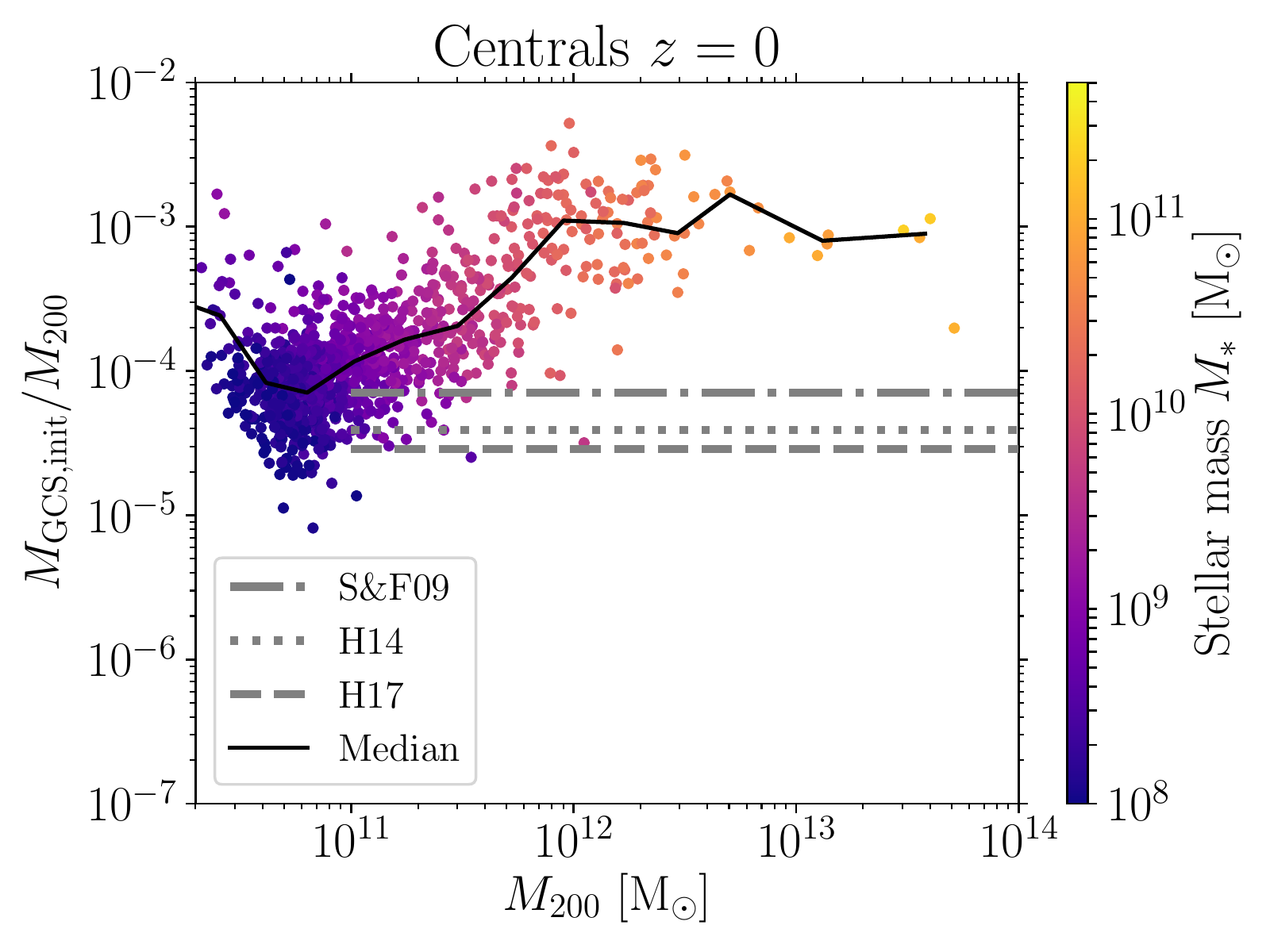}
\caption{`Initial' $\eta$ versus M$_{\rm 200}$, obtained using the initial masses of the clusters.  The flat (linear) relation is still evident above $5\times10^{11}$~\msun, which is due mainly to the hierarchical assembly of galaxies. However, the normalisation increases by a factor of 15 at high halo masses and by a factor of 10 at low masses, which causes this relation to be somewhat steeper than the one including cluster dissolution. }
\label{fig:eta_initial}
\end{figure}

We also investigate the role of cluster disruption on the normalisation of the \mgc-\mhalo\ relation.  In Fig.~\ref{fig:eta_initial}, we show the relation with the present day halo mass but now for the initial masses of GCs (i.e., before mass loss through stellar evolution and cluster disruption).  The overall shape of the distribution is similar, with the turn-down at the same halo mass, but now the normalisation is a factor of $\sim10$ higher than observed.  Including GC mass loss (and full disruption) moves the high mass end onto the observed relation and flattens the overall relation.  This is due to GC disruption being more efficient in higher mass galaxies than less massive counterparts.  We conclude that cluster disruption plays a strong role in setting the normalisation of relation \citep[as predicted by][]{Kruijssen_15}, and models that neglect mass loss (or only include secular mass loss and neglect tidal shocks) will either over-predict $\eta$ at $z=0$ or, if calibrated at $z=0$, will predict less evolution with redshift than models that do include cluster disruption.


\subsection{Cause and Implications of the Downturn at low mass}

\defcitealias{Forbes_et_al_18}{F18}

An important feature visible in the top-left panel (fiducial model) of Fig.~\ref{fig:eta} is that below a halo mass of $5 \times 10^{11}~\msun$ the \mgc-\mhalo\ relation begins to deviate from being linear.  As discussed in Section~\ref{sec:intro}, this appears to be in tension with observations that have not detected such a downturn to date 
\citet[hereafter \citetalias{Forbes_et_al_18}]{Forbes_et_al_18}
However, this downturn is a consistent prediction of models for the GC population, as it is also seen in the models of \citet{El-Badry_et_al_19} and \citet{Choksi_and_Gnedin_19}.  In particular, \citet{El-Badry_et_al_19} find that this downturn is due to the lower gas surface densities and higher mass-loss rates of lower mass galaxies. As a result, both their fiducial model and ours produce fewer clusters at low gas surface densities.

We can demonstrate this in our model by considering the results for models with different cluster formation physics.  In particular, Fig.~\ref{fig:eta} shows that fixing the CFE to a constant value results in a flatter distribution, although there is still a downturn at low halo masses.  In order to retain a flat distribution at lower values of \mhalo\ we would need to increase the CFE towards lower mass galaxies, which would be contrary to the adopted model and also inconsistent with the observed CFEs in nearby dwarf galaxies in the Universe today \citep[e.g.,][]{Cook_et_al_12, Adamo_Bastian_18}.

Even when adopting a constant CFE, there is still a downturn at low galaxy masses.  This is driven by the shape of the \mstar-\mhalo\ relation of the parent galaxies, as this relation is rapidly changing below $\mhalo \sim 5 \times10^{11}~\msun$ \citep[e.g.,][]{Behroozi_et_al_13,Moster_et_al_13,Schaye_et_al_15}.  Adopting a shallower \mstar-\mhalo\ relation than that found in the EAGLE simulations could, in principle, flatten the resulting \mgc-\mhalo\ relation (see Section \ref{sec:inference}).

In the right panels of Fig.~\ref{fig:eta}, we show the total amount of mass in GCs compared to the stellar mass (\mstar) of each galaxy.  Additionally, the observational results from \citet{Peng_et_al_08} are shown as a dashed line.  In this space, we find that our simulations follow the observations relatively well, and do not display any distinctive feature at the low stellar mass end, which is consistent with the observations.  In our simulations, cluster formation and evolution is more closely tied to the formation of the stellar mass of the host galaxy than the dark matter halo or virial mass, and the ability of the simulations to reproduce the observed trend between GCs and the stellar mass suggests that the models are capturing much of the essential physics.  The possible failure in reproducing the trend at low halo masses between the halo and GC system masses may rather reflect differences in relating \mhalo\ to \mstar, as \mstar\ is the more directly measurable quantity.  This is discussed in more depth in Section~\ref{sec:stellar_relations}.

\subsection{Centrals versus satellites}
\label{sec:satellites}

\begin{figure}
\centering
\includegraphics[width=8cm]{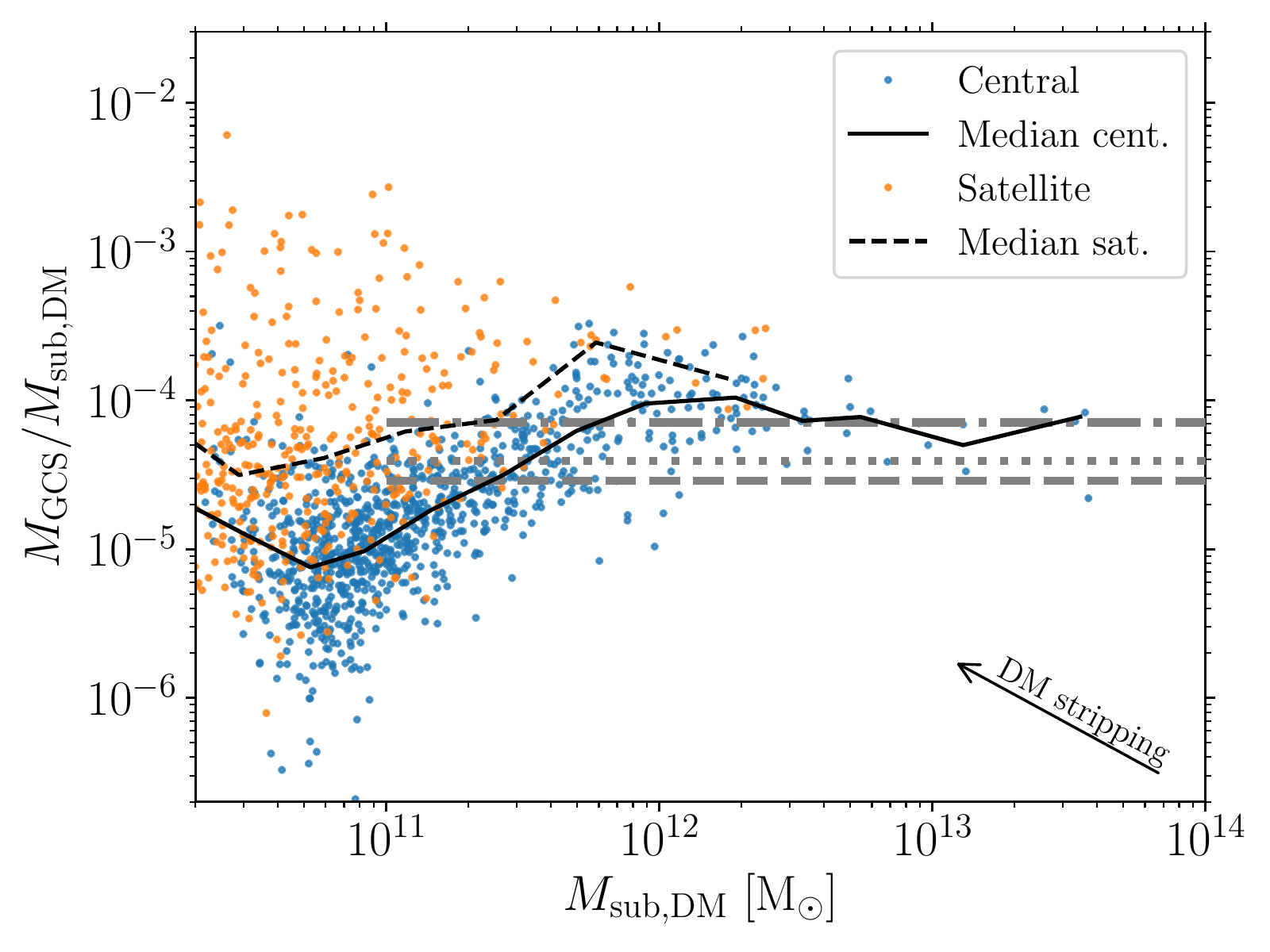}
\caption{ The \mgc-\mhalo\ relation for central (blue) and satellite (orange) galaxies.  We also show the median relations for central (solid lines) and satellite (dashed lines).  Stripping of the dark matter halo (which happens much earlier than stripping of the stellar or GC component of the galaxies) causes the satellite population to scatter up and to the left in the figure.  The use of a one-to-one conversion of stellar to halo mass based on central galaxies will lead to the over-estimation of the actual halo mass for satellite galaxies.}
\label{fig:satellites}
\end{figure}

So far, we have focused our attention on central galaxies, as the majority of observational work either focused directly on centrals or uses \mstar-\mhalo\ relations that are built from observations of central galaxies \citepalias[e.g.,][]{Harris_et_al_17}.  In Fig.~\ref{fig:satellites}, we show the \mgc-\mhalo\ relation for our sample of central galaxies (blue dots) as well as a sample of satellite galaxies (orange dots).  We also show the running medians of each distribution of centrals and satellites as a solid and dashed black line, respectively.  

From Fig.~\ref{fig:satellites}, it is clear that at a fixed halo mass, satellite galaxies have a larger $\eta=\mgc/\mhalo$ ratio, which is particularly noteworthy for halo masses less than $\sim5\times10^{11}$~\msun.  This is due to the tidal stripping of the satellite galaxy's dark matter halo by their central galaxy, causing the affected satellites to move to the upper left of the figure. This differential effect of tidal stripping occurs because the dark matter haloes of galaxies are spatially more extended than their GC populations.

\citetalias{Forbes_et_al_18} have investigated the \mgc-\mhalo\ relation in a sample of nearby galaxies, using rotation curves to measure the dynamical mass of the galaxy halo.  These authors report that the \mgc-\mhalo\ relation remains linear (i.e., does not show a pronounced downturn) down to a halo mass of $\sim10^8$~\msun \footnote{Although if galaxies without detected GCs are included the running mean/median would show a significant downturn below a halo mass of $\sim10^{10}$~\msun.}. However, the \citetalias{Forbes_et_al_18} sample contains a number of satellite galaxies (in particular at low masses). As shown in Fig.~\ref{fig:satellites}, this is expected to lead to a flatter \mgc-\mhalo\ relation. We will return to this point in \S~\ref{sec:inference}.

\subsection{Redshift dependence}

\citet{Choksi_and_Gnedin_19} have used their model for GC formation and evolution within galaxies to trace the evolution of the \mgc-\mhalo\ relation from a redshift of 10 until today.  In their model, the shape of the relation is already set at $z=10$ and the normalisation evolves slowly from $z=10$ to $z=3$ (by a factor of $\sim2$).  Between $z=3$ and $z=0$, the shape of the relation still remains the same, but the evolution of the normalisation accelerates as it decreases by a factor of $\sim10$.  The authors find that this rapid change is driven primarily by the build up of halo mass from $z=3$ to $z=0$, as the GC system mass remains largely constant, reflecting a rough balance between dynamical GC disruption and late GC formation and accretion.

\begin{figure*}
\centering
\includegraphics[width=16cm]{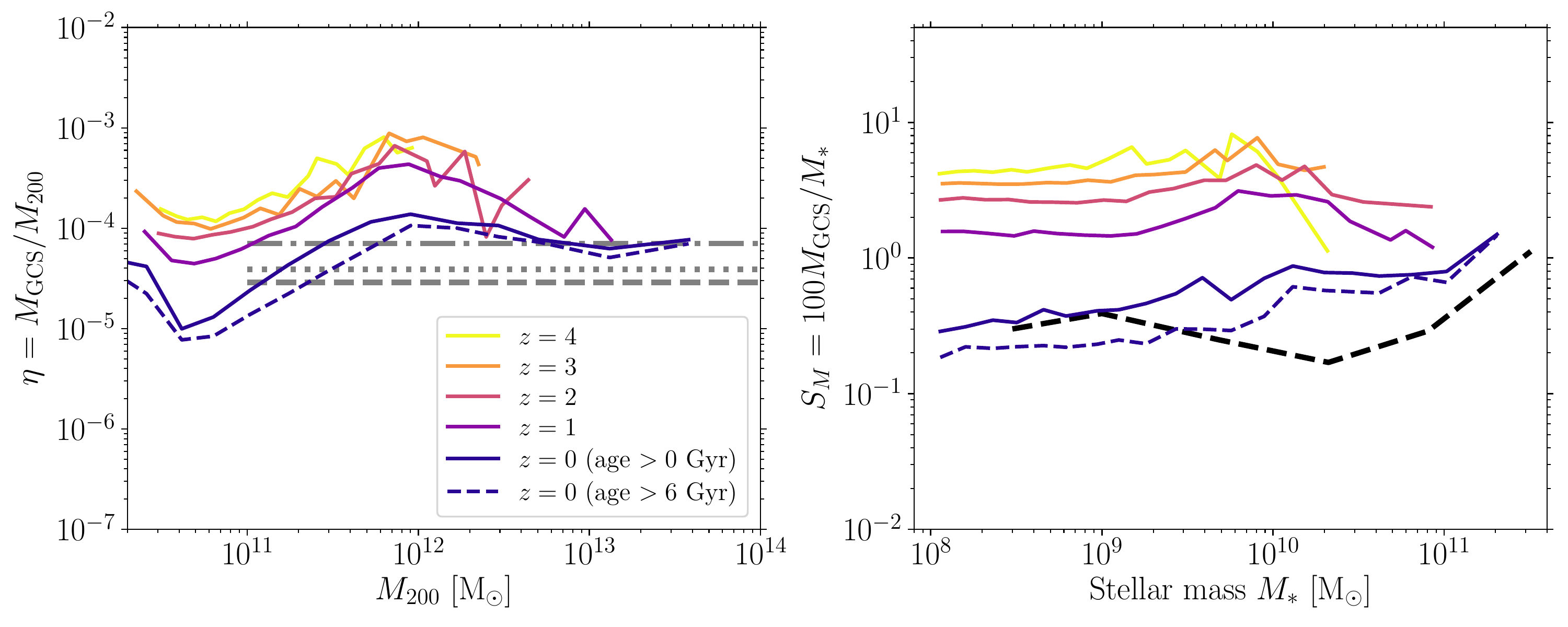}
\caption{The median \mgc-\mhalo\ relation (left panel) and the median \sm-\mstar\ relation (right panel) for central galaxies at different redshifts, for the fiducial model.  The \mgc-\mhalo\ relation is set early on (at $z>4$) and exhibits little evolution to $z\sim1$, after which the relation evolves considerably to $z=0$.  We also show the $z=0$ relation for two different GC age cuts, $>6$~Gyr (dashed line) and for all ages.  Their similarity argues for a minor contribution of late-time GC formation to this relation.  A similar conclusion is reached for the \sm-\mstar\ relation (right panel), for which we also show the observed relation from \citet{Peng_et_al_08} (long dashed black line).}
\label{fig:zevo}
\end{figure*}

In the left panel of Fig.~\ref{fig:zevo}, we show the median \mgc-\mhalo\ relation as a function of redshift for our model galaxies.  Like \citet{Choksi_and_Gnedin_19}, we find that the overall shape of the relation is set at early times ($z>4$ in our simulations) and evolves slowly to a redshift of $\sim1$, followed by a rapid drop to $z=0$.  Quantitatively, we predict a drop of a factor of $\sim7-10$ from $z=2$ to $z=0$, in good agreement with the results of \citet{Choksi_and_Gnedin_19}.  For $z=0$, we show the results for two different GC age cuts. One line shows all GCs (solid purple line), whereas the other only includes GCs older than $6$~Gyr at $z=0$ (dashed purple line).  The similarity of the two relations, especially at large halo masses, suggests that late-time GC formation is a relatively minor driver of the normalisation of this relation.

\begin{figure*}
\centering
\includegraphics[width=16cm]{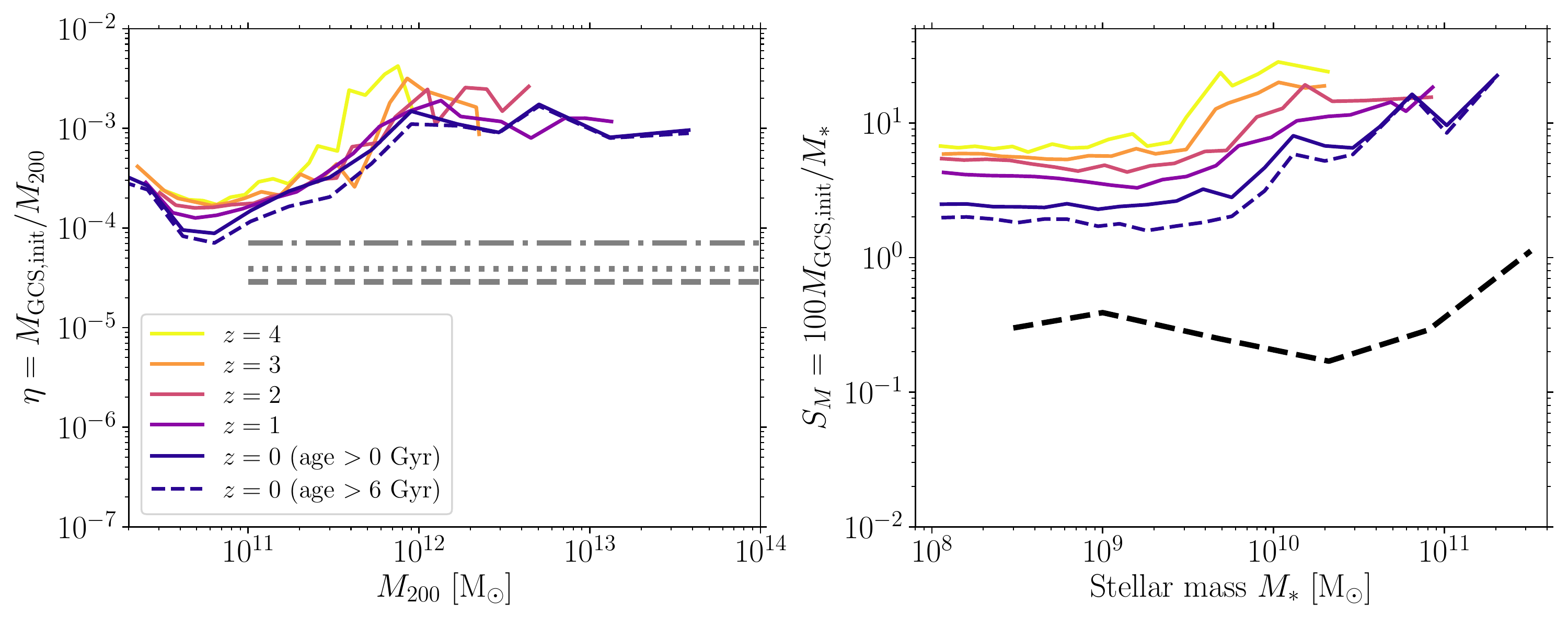}
\caption{The same as Fig.~\ref{fig:zevo}, but calculating \mgc\ using the initial cluster masses (\mhalo\ is the same for each galaxy as shown previously). The similarity in the \mgc-\mhalo\ relation (when initial cluster mass is used) at all redshifts shows that cluster dissolution is the driving cause behind the \mgc-\mhalo\ relation's evolution with redshift. For the \sm-\mstar\ relation, the evolution with redshift (when initial cluster masses are used) is a reflection of formation bias, where galaxies that reach a certain stellar mass earlier did so through a higher mean SFR than galaxies that reach that mass later.  More intense bursts of star-formation leads to a higher fraction of the stellar mass in GCs.}
\label{fig:zevo_initial}
\end{figure*}

In order to find the origin of the redshift dependence of the \mgc-\mhalo\ relation, in Fig.~\ref{fig:zevo_initial} we show the same relations, but now using the cluster initial masses. When the initial masses are used, we see little evolution with redshift, hence it appears that cluster disruption is the driving cause of the redshift evolution in the simulations \citep[as predicted by][]{Kruijssen_15}\footnote{While the E-MOSAICS simulations likely under-predict the rate of cluster dissolution \citep[see][]{Kruijssen_et_al_19a}, by concentrating our analysis on the high-mass end of the GC mass function the effect should be minimised.}. Fig.~\ref{fig:zevo_initial} also shows the effects of formation bias, i.e., that galaxies that form early through intense bursts of star and cluster formation have a higher fraction of their mass in GCs than galaxies that grow to the same mass more gradually through cosmic history \citep[e.g.,][]{Mistani_et_al_16,Kruijssen_et_al_19a}. For example, a galaxy with $\mstar = 10^9$~\msun at $z=4$ has nearly three times the mass in GCs than the same mass galaxy at $z=0$.  Formation bias within the E-MOSAICS simulations will be explored in more detail in Crain et al. (in prep.).

In the right panel of Fig.~\ref{fig:zevo} we show the \sm\ versus \mstar\ relation from our simulations at different redshifts.  Additionally, the black dashed line shows the \citet{Peng_et_al_08} relation for observed galaxies at $z=0$.  We find a similar, albeit more gradual evolution of the relation as the \mgc-\mhalo\ relation, suggesting that the stellar mass of the host galaxy is also growing from $z=2$ to $z=0$ without an appreciable change in GC system mass.

In the context of Figs.~\ref{fig:zevo} and~\ref{fig:zevo_initial}, we note that \citet{El-Badry_et_al_19} additionally investigated a ``pathological" model, in which the mass in GCs is uncorrelated to that of the mass of the halo.  In this type of model, reproducing the $z=0$ \mgc-\mhalo\ relation requires the opposite redshift evolution of that seen in our models, such that $\eta$ at fixed halo mass is lower at higher redshift. Future observations with JWST or the E-ELT of GC populations as a function of redshift may be able to test these predictions.


By contrast, the fiducial model of \citet{El-Badry_et_al_19} shows a much milder evolution with redshift than both the present study and that of \citet{Choksi_and_Gnedin_19}.  For their fiducial model, \citeauthor{El-Badry_et_al_19} find that the normalisation of the \mgc-\mhalo\ relation changes by only a factor of $\sim3$ between $z=3$ and $z=0$.  This weaker evolution, in contrast to the model presented here, may be a result of the lack of cluster disruption in the \citeauthor{El-Badry_et_al_19} simulations.  Hence, in order to reproduce the $z=0$ relation with little or no disruption, it is necessary to have less evolution of the relation towards higher redshift (see \S~\ref{sec:normalisation}). In this context, we note that the overall GC formation history in the model of \citet{El-Badry_et_al_19} is similar to that of E-MOSAICS \citep{Reina-Campos_et_al_19}.  



\subsection{Linking GC Populations, Stellar Content, and Halo Mass of Galaxies}
\label{sec:inference}

In the majority of observational work on the \mgc-\mhalo\ relation, the dark matter content of each galaxy was not measured directly, but rather was inferred from scaling relations between the stellar mass and the dark halo mass obtained from abundance matching or weak lensing surveys.  While the \mstar - \mhalo\ relation is relatively well constrained at high halo masses ($\mhalo > 10^{11}$~\msun), below this mass differences between studies can amount to two orders of magnitude or more in the stellar mass at fixed halo mass.  This uncertainty can have a large and important impact on the inferred \mgc - \mhalo\ relation when translating from the observational plane of \mgc - \mstar.

Two exceptions to the above are the studies of \citet{forbes_et_al_16} and \citetalias{Forbes_et_al_18}, who used dynamical tracers to estimate the halo mass (although correction factors for each galaxy needed to be applied to extrapolate from the outermost dynamical measurement to \m200).  In particular, \citetalias{Forbes_et_al_18} attempted to trace the \mgc - \mhalo\ relation down to low halo masses to see whether the observed linear relation at high masses continues, or if a break appears in the relation.  As discussed in the introduction and \S~\ref{sec:satellites}, the authors report that the relation (at least for galaxies that host GCs) continues to be linear down to at least $\mhalo \sim 10^{8}$~\msun.  This is in apparent contradiction to the predicted \mgc - \mhalo\ relation found in our simulations, which features a notable downturn at $\mhalo=5\times10^{11}$~\msun. 

For each of the galaxies in their sample, \citetalias{Forbes_et_al_18} measured \mgc, \mstar, and \mhalo.  As noted by the authors, their derived \mstar - \mhalo\ relation is strongly inconsistent with the relation inferred through abundance matching, as well as empirical studies and simulations.  Specifically, the results of \citetalias{Forbes_et_al_18} imply stellar masses $>1-2$ orders of magnitude larger\footnote{This is not due to the uncertainty in $\mstar$, but is instead caused by the large uncertainty in the inference of $\mhalo$ using dynamical models.} (at a halo mass of $10^{9}$~\msun) than more canonical relations.

\begin{figure}
\centering
\includegraphics[width=8cm]{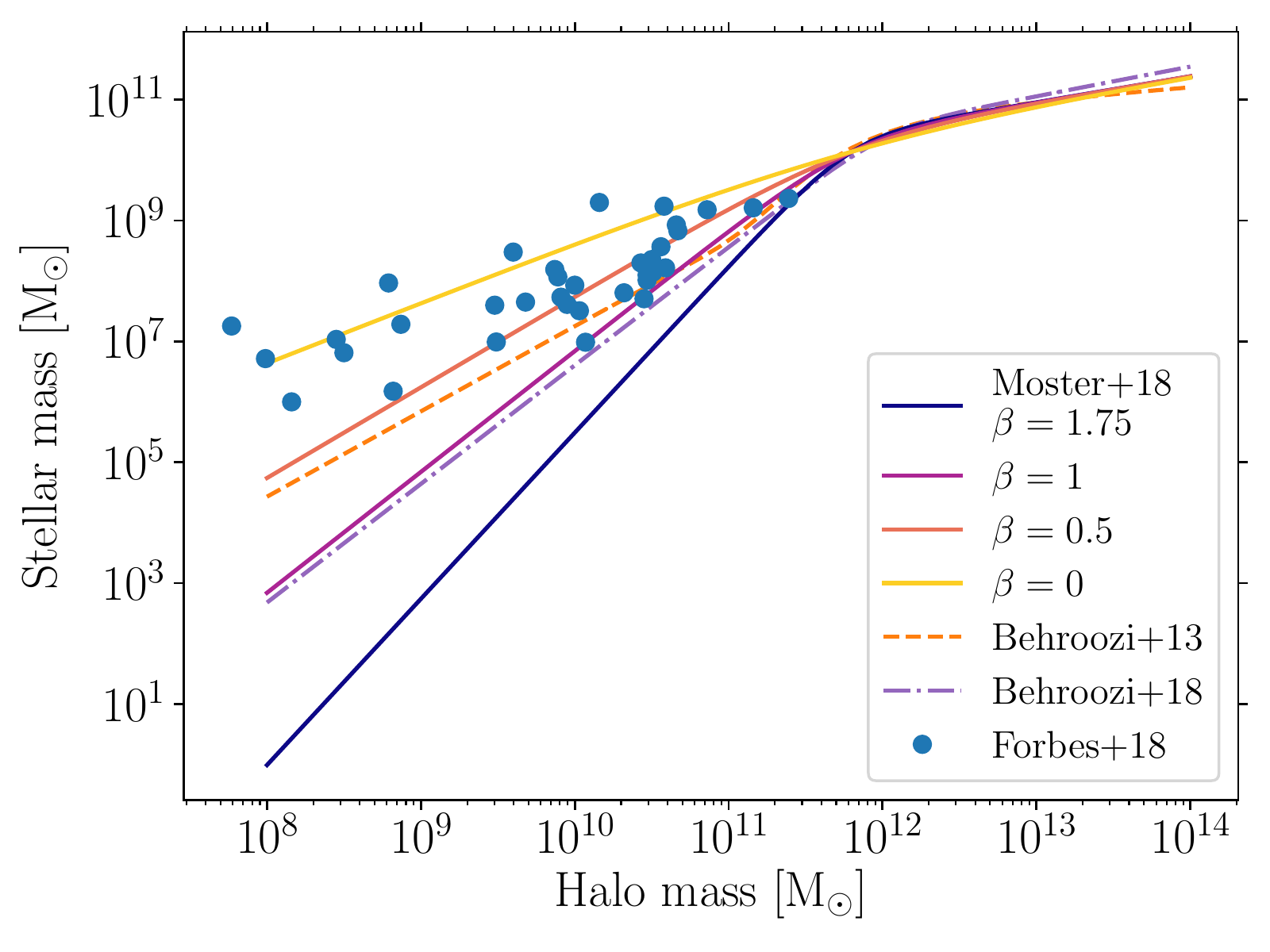}
\includegraphics[width=8cm]{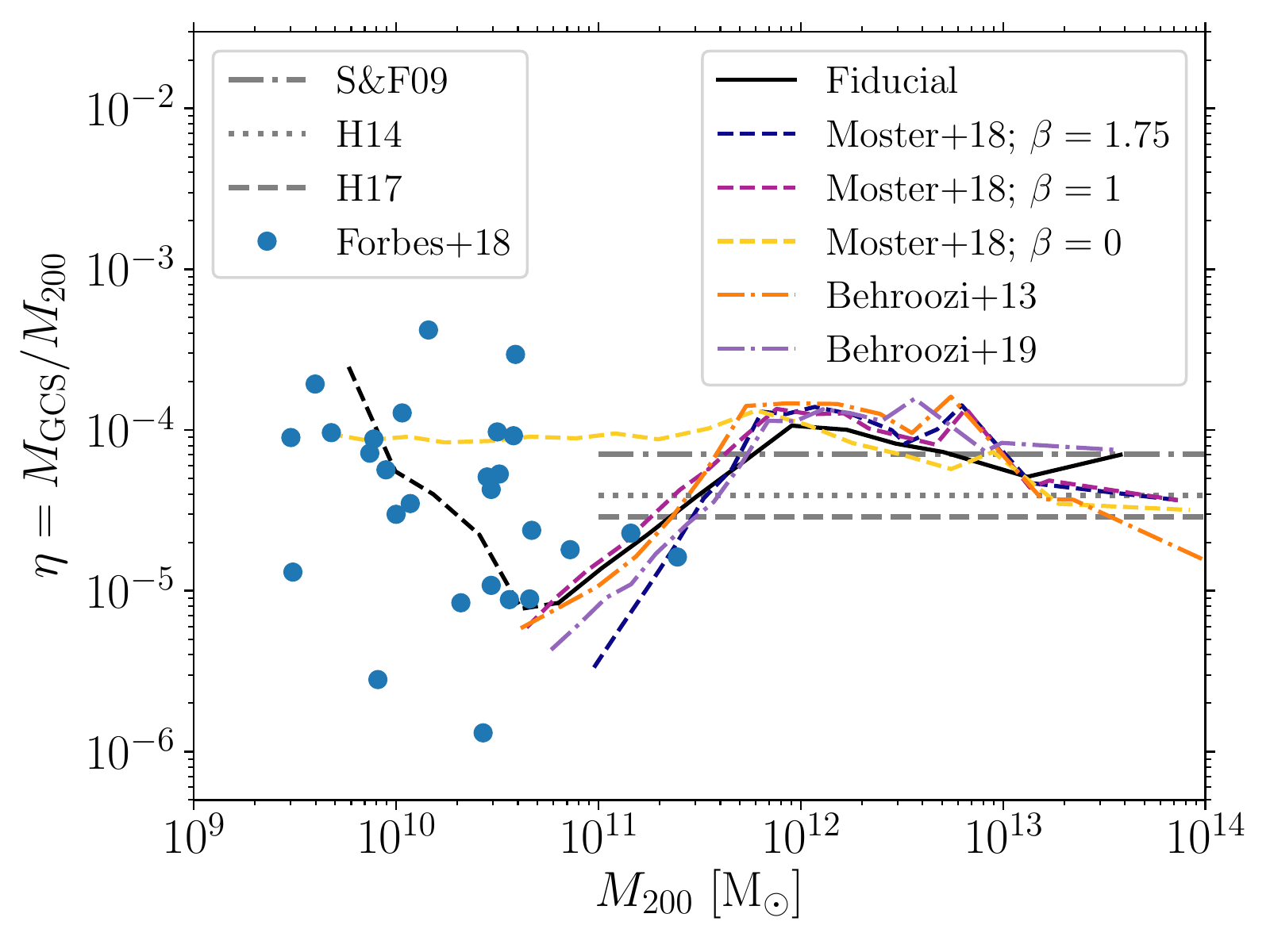}
\caption{
{\bf Top panel:} The \mstar - \mhalo\ relation of galaxies from a variety of sources.  For the \citet{moster_et_al_18} relation we show the result of varying  the slope of the low mass end.  The measurements (for galaxies that contain GCs) from \citetalias{Forbes_et_al_18} are shown as solid circles, and lie significantly above the more canonical relations in the literature. {\bf Bottom panel:}  The resulting \mgc - \mhalo\ relation for our fiducial model when translating our \mgc - \mstar\ relation using the stellar/halo mass relations from the top panel.  Depending on the adopted \mstar - \mhalo\ relation, the \mgc - \mhalo\ relation may display a distinct and steep downturn or even remain flat (this can only be achieved by setting $\beta=0$, implying $\mstar\propto\mhalo$).  The dashed black line shows our fiducial model below a halo mass of $5\times10^{10}$~\msun), which begins to increase due to our imposed stellar mass cut ($10^{8}~\msun$, see text for details).}

\label{fig:smhm}
\end{figure}

In the top panel of Fig.~\ref{fig:smhm}, we show the \mstar - \mhalo\ relation as measured by \citet{Behroozi_et_al_13,Behroozi_et_al_19}.  Additionally, we show the \citet{moster_et_al_18} relation who adopt a double power-law formulation, with a low mass slope, $\beta$, for which we show values in the range $\beta=0-1.75$.  The value preferred by \citet{moster_et_al_18} is $\beta=1.75$ at redshift 0 (the blue, lowest, line).  We also show the individual data points from \citetalias{Forbes_et_al_18}, which below a halo mass of $\sim10^{11}$~\msun\ are significantly above the canonical relations.  If parametrized in the same way as done in \citet{moster_et_al_18}, the Forbes et al. data would imply $ 0 < \beta < 1$, significantly shallower than other studies.

These differences have important implications when translating into the \mgc - \mhalo\ plane.  In the bottom panel of Fig.~\ref{fig:smhm}, we show the resulting \mgc - \mhalo\ relation, derived from our fiducial model using the median \sm - \mstar\ relation from the top right panel of Fig.~\ref{fig:eta} and the different \mstar - \mhalo\ relations shown in the upper panel.  The solid line shows the relation directly measured in our fiducial simulation.  Other lines show the resulting relation when adopting different \mstar - \mhalo\ relations shown in the top panel.  The solid points are the measurements for the galaxies from \citetalias{Forbes_et_al_18} shown in the top panel.  

The dashed black line in the bottom panel of Fig.~\ref{fig:smhm} shows the results of our fiducial model where it begins to increase due to our imposed stellar mass cut ($10^{8}$~\msun - corresponding to a halo mass of $5\times10^{10}$~\msun).  Below this halo mass (due to the fixed stellar mass cut), we are only sensitive to the most extreme galaxies in terms of the $\mstar / \mhalo\ $ ratio.  These galaxies are part of a biased sub-population that tend to have larger GC populations than other galaxies at comparable halo masses, resulting in elevated $\eta$ values.  This may also be affecting observational work and can only be overcome through complete, volume limited, surveys.

The results in Fig.~\ref{fig:smhm} show that for a given, observed \sm - \mstar\ relation the resulting inferred \mgc - \mhalo\ relation can show a distinct downturn or even be flat, depending on the adopted \mstar - \mhalo\ relation.  Hence, uncertainties in the shape of the \mstar - \mhalo\ relation at low masses, as well as inconsistencies between observations and simulations, mean that, at present, the low mass end of the \mgc - \mhalo\ relation cannot be used to place strong constraints on our understanding of GC formation and/or evolution.


We further investigate the importance of the \mstar - \mhalo\ by looking at two observational samples that use \mgc\ and \mstar.  The first is the median relation from \citet{Peng_et_al_08}, which is shown as a dashed line in the right panels of Fig.~\ref{fig:eta}.  The second is the \citetalias{Forbes_et_al_18} sample, where we adopt their measured \mgc\ and \mstar\, i.e. we do not use their inferred \mhalo\ values.  Using these datasets we then apply two commonly adopted \mstar-\mhalo\ relations from the literature, namely those of \citet[][with $\beta=1.75$]{moster_et_al_18} and \citet{Behroozi_et_al_13}.

\begin{figure}
\centering
\includegraphics[width=8cm]{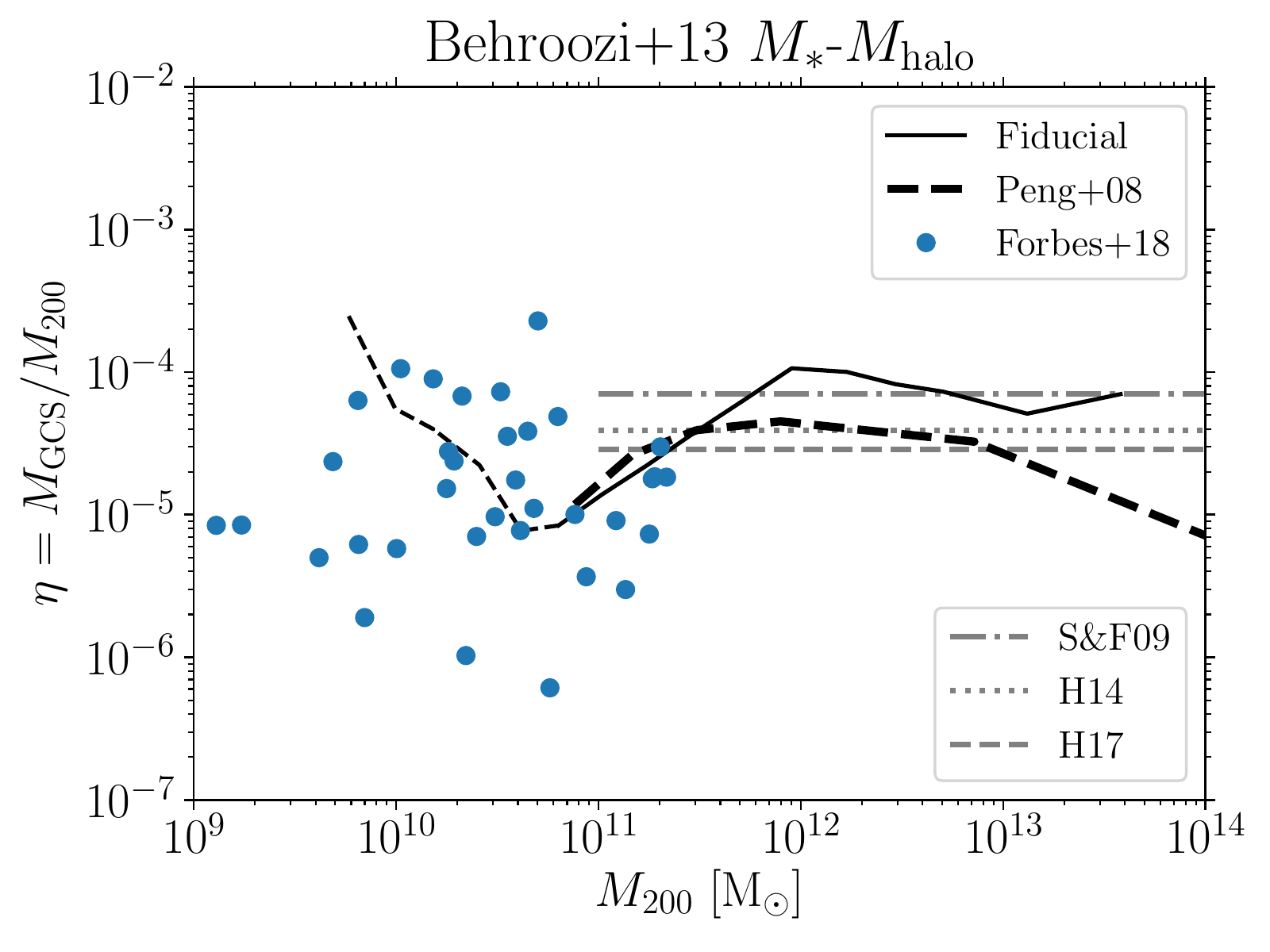}
\includegraphics[width=8cm]{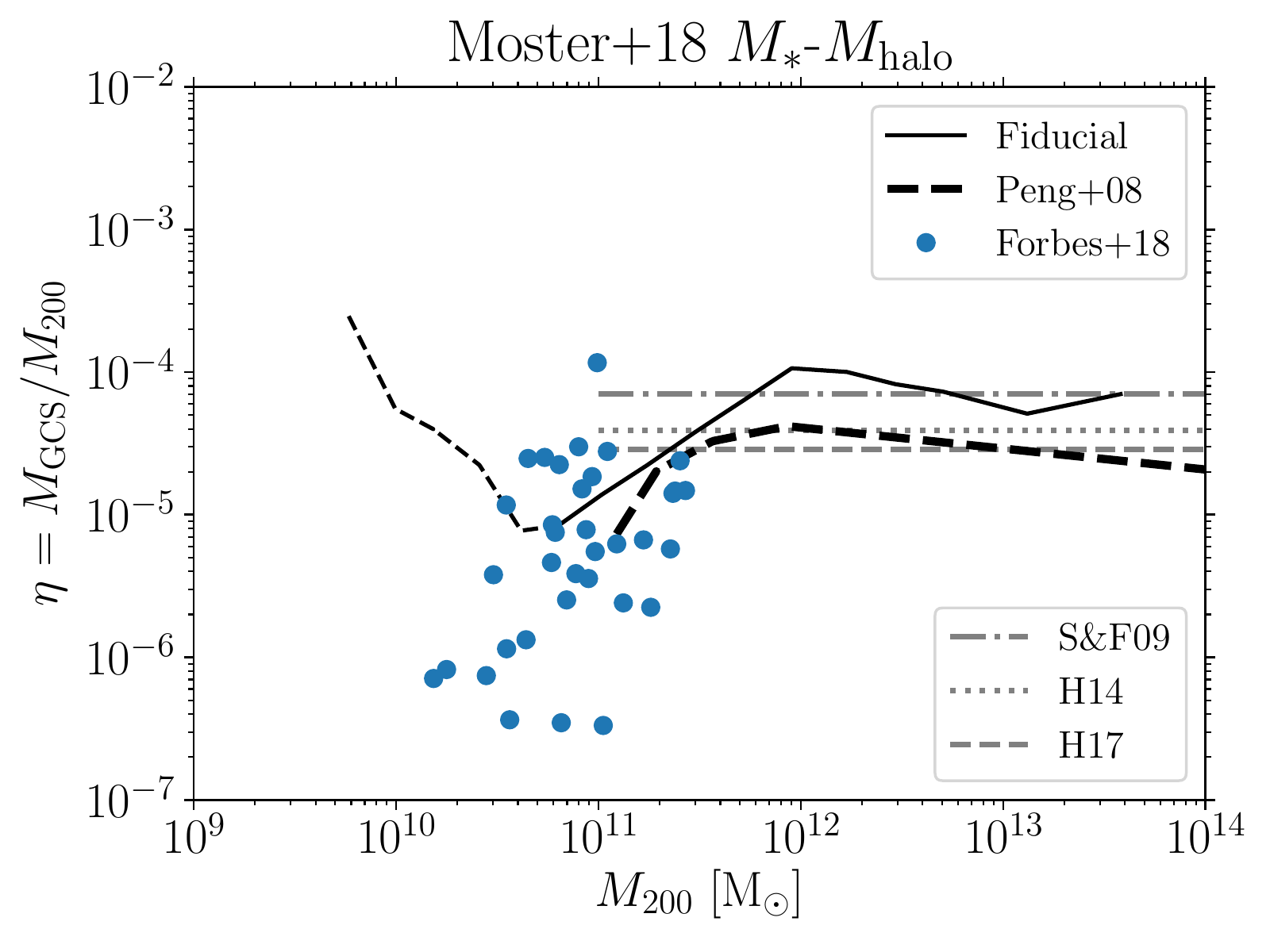}
\caption{ 
{\bf Top panel:} The \mgc - \mhalo\ relation inferred from the observed \sm - \mstar\ relation from \citet{Peng_et_al_08} (dashed line), and \citetalias{Forbes_et_al_18} (filled circles) by adopting the \citet{Behroozi_et_al_13} \mstar - \mhalo\ relation. Additionally, we show the results from our fiducial model as a solid line. {\bf Bottom panel:} The same as the top figure but now adopting the \citet{moster_et_al_18} \mstar - \mhalo\ relation.  Note that in both cases the observations show a downturn at or near the same halo mass ($\sim5\times10^{11}$~\msun) as our simulations.  From this we conclude that, given the uncertainty in the \mstar - \mhalo\ relation, the observed \mgc - \mhalo\ relation is consistent with the relation found in the E-MOSAICS simulations.}
\label{fig:behroozi}
\end{figure}

The resulting, observationally-inferred \mgc - \mhalo\ relations are shown in Fig.~\ref{fig:behroozi}.  In the upper panel we adopt the \citet{Behroozi_et_al_13} relation while in the lower panel we adopt the \citet{moster_et_al_18} relation.  Rather than appearing as a flat continuation from the results obtained for higher mass galaxies (shown as horizontal lines), both the \citet{Peng_et_al_08} and \citetalias{Forbes_et_al_18} data show a clear downturn at or near the same mass found in our simulations (i.e., $\mhalo \sim 5\times10^{11}$~\msun).  Again, we note that the simulations of \citet{El-Badry_et_al_19} and \citet{Choksi_and_Gnedin_19} found a downturn at a similar mass.

{\em We conclude that the observations of GC systems to date are  consistent with the downturn in the \mgc - \mhalo\ relation predicted in our model (within the large uncertainties in halo masses of low mass galaxies).  This downturn is not driven by the physics of GC formation or evolution, but rather is a simple reflection of the \mstar - \mhalo\ relation as a function of halo mass.  Hence, until the \mstar - \mhalo\ relation at the low mass end is pinned down, and the origin of the discrepancy between the relations seen in observations and simulations is known, the shape of the \mgc - \mhalo\ relation at the low mass end will tell us little about how GCs relate to their host galaxies or when and where they formed.}

\subsection{\sm-\mstar\ relation}
\label{sec:stellar_relations}

As noted above, in most observational studies, the halo mass of the host galaxy is not directly measured, but rather it is inferred, often through the assumption of a single relation to translate \mstar\ to \mhalo (notable exceptions to this are \citet{forbes_et_al_16} and \citetalias{Forbes_et_al_18} discussed above).  Hence, the \mgc-\mhalo\ relation is not a direct relation, but contains two components.  The first is the \mgc-\mstar\ relation and the second is the \mstar-\mhalo\ relation.  This double dependence can make it difficult to isolate the physical effects driving any relation between GCs and their host galactic dark matter halo.  As such, it is more insightful to investigate relations between quantities that can be directly observed.  

Additionally, we argue that the use of the number of GCs (\ngc), which is adopted in some observational studies as a substitute for \mgc, is not an ideal quantity as it can be strongly affected by observational selection effects.  For example, studies of local group galaxies often include GCs with masses below $10^4$~\msun\ \citepalias[e.g.,][]{Forbes_et_al_18} while studies of more distant systems are limited to clusters above a few $10^5$~\msun\ \citepalias[e.g.,][]{Harris_et_al_17}.   Similarly, quantities like the specific frequency (S$_{\rm N}$, i.e.\ the number of clusters per unit galaxy luminosity) are known to exhibit extreme stochasticity in the limit of low numbers of GCs, such that a single GC (independent of its mass) in a low-luminosity galaxy results in a high value of S$_{\rm N}$, producing a statistical rather than a physically-driven upturn.

For these reasons, we argue that the most physically-driven estimator of the richness of a GC population is the total mass in GC per unit galaxy stellar mass, $\sm = 100 \times \mgc / \mstar$.  While this quantity has historicaly been less commonly used in observational studies relative to S$_{\rm N}$ or $\eta$, it has been explored in some studies \citep[e.g.,][]{Georgiev_et_al_10}.

In the right-hand panels of Fig.~\ref{fig:eta}, we show \sm\ versus \mstar\ for our simulations.  For the fiducial model, we find the \sm-\mstar\ relation to be nearly flat (and hence the \mgc-\mstar\ relation to be nearly linear), with a slight upturn at high galaxy masses.  We also show the observed relation from \citet{Peng_et_al_08}, which is also quite flat, opposed to the `U'-shaped relation observed between the specific frequency and galaxy stellar mass or luminosity.  The normalisation of our fiducial model is also quite close to that of the \citeauthor{Peng_et_al_08} observational sample.

As discussed in \citet{Peng_et_al_08} (highlighted in their Figure 6) the "U" shape observed in the relation between the specific frequency (S$_{N}$) and galaxy luminosity (M$_{z}$) corresponds to a much flatter relation between specific mass (S$_{M}$) and galaxy stellar mass (\mstar) due to the mean GC luminosity decreasing with galaxy luminosity.  We chose to compare our models with the observed S$_{M}$-\mstar\ relation, rather than in specific frequency due the sensitivity of S$_{N}$ on the shape of the GC mass function at low masses.  As E-MOSAICS underestimates cluster disruption, in particular at lower masses, the total mass in GCs is expected to be more robust than the number of GCs.

\citetalias{Harris_et_al_17} found an upward kink in the \mgc-\mstar\ relation at a galaxy stellar mass of $\sim10^{10}$~\msun, implying that more massive galaxies have a higher fraction of their stellar mass in GCs than lower mass galaxies.  Our fiducial simulations reproduce this upward kink and do so at a similar galaxy mass (see the top-right panel in Fig.~\ref{fig:eta}).  Looking at the `\mcstar\ only' model, no such upward trend is seen, suggesting that the cause of the trend is due to a higher CFE in higher-mass galaxies.  We find that the upturn is also present when using the initial GC masses at high redshift (see Fig.~\ref{fig:zevo_initial}), lending further credence to the CFE interpretation.


We note that the low-mass galaxies used in \citetalias{Forbes_et_al_18} are consistent with the  \sm - \mstar\ relation from \citet{Peng_et_al_08}, hence are consistent with the results of the simulations presented here.  Again, this argues that the differences between the observed and simulated \mgc-\mhalo\ relations are driven by differences in the underlying \mstar-\mhalo\ relation, and are not directly related to the GC population properties.

Looking at the other models that differ in their input GC formation physics in Fig.~\ref{fig:eta}, we find that they also result in changes to the \mgc-\mstar\ relation.  If we adopt a model with a pure-power law mass function, but allow the CFE to vary, the resulting simulations do not reproduce the observed normalisation and also display a trend with galaxy mass that is inconsistent with the observations.  Alternatively, for a fixed CFE ($10$\%), but variable \mcstar, the simulations reproduce the normalisation of the \sm - \mstar\ relation, but fails to reproduce the observed upturn at high halo masses.


\section{Discussion and Conclusions}
\label{sec:discussion}

We have analysed the \mgc-\mhalo\ relation realised in a large cosmological volume ($34.4$~cMpc on a side) simulated with the E-MOSAICS model, in order to establish the origin of the relation and its evolution over cosmic time, and to explore its sensitivity to the input physics of GC formation and disruption\footnote{We note that the fiducial E-MOSAICS model was originally designed and tested using a suite of 10~Milky Way-mass cosmological zoom-in simulations. The \mgc-\mhalo\ relations presented here apply the identical model to a complete cosmological volume, and the predicted trends with galaxy mass thus represent a genuine prediction.} Within the simulated volume, we follow the formation and evolution of the GC population in 1707 galaxies with stellar masses above $10^8$~\msun\ at $z=0$.  In agreement with recent other models, such as \citet{Kruijssen_15}, \citet{El-Badry_et_al_19} and \citet{Choksi_and_Gnedin_19}, we find a linear \mgc-\mhalo\ relation above a halo (M$_{\rm 200}$) mass of $\sim5\times10^{11}$~\msun.  This is driven essentially by the process of hierarchical galaxy growth through the central limit theorem.

Above $\sim5\times10^{11}~\msun$, i.e.\ in the linear regime, we have investigated the effects that control the normalisation, i.e.\ why $\eta\equiv\mgc/\mhalo \sim 5 \times10^{-5}$.  Unlike most previous numerical models that had to set certain free parameters in order to reproduce this value, we are able to adjust the input physics directly to uncover the origin of the normalisation.  We find that not including an environmentally dependent term for the cluster formation efficiency ($\Gamma$) or the mass function exponential truncation (\mcstar) leads to a systematic offset from the observations.  Our fiducial model reproduces the observed value without adjusting any parameters.  Likewise, we have investigated the role of cluster disruption in setting the normalisation and have found that it plays a dominant role \citep[as predicted by][]{Kruijssen_15}.  By not including GC disruption (which is dominated by tidal shocks) the normalisation is off by a factor of $\sim10$.

Our simulations predict a downturn in the \mgc-\mhalo\ relation below $\sim5\times10^{11}~\msun$.  
By changing the input physics of cluster formation, specifically $\Gamma$ and \mcstar, we show that the formation physics are unlikely to be the underlying cause of the downturn.  Similarly, by looking at the simulation results with and without cluster mass loss (through stellar evolution and the loss of stars from internal dynamics and external tidal effects), we show that cluster disruption is also unlikely to be the origin of the downturn.
Instead, the cause of the downturn is the underlying relation between the stellar mass (\mstar) and halo mass (\mhalo) of galaxies.  By applying a range of \mstar - \mhalo\ relations from the literature to observed samples of GC systems and their host galaxies, we show that the observations presented to date are consistent with the predicted downturn from our simulations.

Additionally, we show that satellite galaxies tend to scatter up and to the left in the $\eta$-\mhalo\ relation, due to the fact that the dark matter halo is more extended than the stellar or GC population of a galaxy, so the dark halo is tidally stripped by the central galaxy before the GCs and stellar mass are.  This results in a flatter $\eta$-\mhalo\ relation, with the downturn being less pronounced.  Along with the adopted \mstar-\mhalo\ relation, this effect, as well as the exclusion of galaxies without GCs, may explain why observational studies to date have not found evidence of the predicted downturn.  In order to definitively establish the form of the \mgc - \mhalo\ relation at low galaxy masses,  volume-complete sample of isolated galaxies that accounts for distance-dependent selection limits would be required.

We note that the results of \citet{burkert20}, which use the number of observed GCs in galaxies as an indicator of their halo masses, is also sensitive to the adopted \mstar-\mhalo\ relation. The use of the \mstar-\mhalo\ relation from the EAGLE simulations (used in this work) would result in a turn-down in the \ngc-\mhalo\ at $\mhalo \sim 5 \times10^{11}$~\msun, analagous to the \mgc-\mhalo\ relation.


Importantly, when exclusively considering the baryonic properties of the galaxies (i.e., using a directly observable quantity such as stellar mass instead of halo mass), namely \sm\ versus \mstar, we find good agreement between the simulations and observations. 
Our fiducial model reproduces the observed, (near-)linear \mgc-\mstar\ relation, including its normalisation.  Since we expect a more direct relation between stellar clusters and the stellar component of their host galaxies, rather than with the dark matter halo, and acknowledging the fact that observations are generally restricted to the visible component of galaxies, we argue that the \mgc-\mstar\ relation is the more physically motivated relation between clusters and their host galaxy.  We find that changing the cluster formation physics (i.e., fixing either the CFE or adopting an ICMF without a truncation mass) results in \mgc-\mstar\ relations clearly at odds with observations.  Hence, this relation can be used as a powerful tool to test cluster population formation physics.

Our fiducial model also reproduces the observed upturn in the \mgc-\mstar\ relation at a galactic stellar mass of $\sim10^{10}~\msun$.  By exploring the behaviour of this relation in the different formation physics simulations, as well as its evolution with redshift (both for present day GC masses and the initial values) we found that the upturn is caused by an increase in the median CFE for higher mass galaxies.

Finally, we predict a strong evolution in both the \mgc-\mhalo\ and \mgc-\mstar\ relations as a function of redshift.  This is due to a combination of (1) GC formation and disruption over extended periods and (2) the continuous build-up of galaxies for both the dark matter and stellar components. Initial observational evidence for this evolution has been identified by inferring the masses of satellite galaxies accreted by the Milky Way over the past 10~Gyr \citep{Kruijssen_et_al_20}. Direct evidence may be obtained by directly observing (young) GC populations at high redshift.

\section*{Acknowledgments}

We thank the referee for their helpful and constructive comments that improved the manuscript.  NB and RAC gratefully acknowledge financial support from the Royal Society (University Research Fellowships).  NB and JP gratefully acknowledge financial support from the European Research Council (ERC-CoG-646928, Multi-Pop). JMDK gratefully acknowledges funding from the Deutsche Forschungsgemeinschaft (DFG, German Research Foundation) through an Emmy Noether Research Group (grant number KR4801/1-1) and the DFG Sachbeihilfe (grant number KR4801/2-1). JMDK, STG, and MRC gratefully acknowledge funding from the European Research Council (ERC) under the European Union's Horizon 2020 research and innovation programme via the ERC Starting Grant MUSTANG (grant agreement number 714907). MRC is supported by a Fellowship from the International Max Planck Research School for Astronomy and Cosmic Physics at the University of Heidelberg (IMPRS-HD). This work used the DiRAC Data Centric system at Durham University, operated by the Institute for Computational Cosmology on behalf of the STFC DiRAC HPC Facility (www.dirac.ac.uk). This equipment was funded by BIS National E-infrastructure cap- ital grant ST/K00042X/1, STFC capital grants ST/H008519/1 and ST/K00087X/1, STFC DiRAC Operations grant ST/K003267/1 and Durham University. DiRAC is part of the National E-Infrastructure. This study also made use of high performance computing facilities at Liverpool John Moores University, partly funded by the Royal Society and LJMUs Faculty of Engineering and Technology.

\section*{Data Availability}
The data underlying this article will be shared on reasonable request to the corresponding author.


\bibliographystyle{mnras}
\bibliography{bibliography}

\begin{thebibliography}{}
\makeatletter
\relax
\def\mn@urlcharsother{\let\do\@makeother \do\$\do\&\do\#\do\^\do\_\do\%\do\~}
\def\mn@doi{\begingroup\mn@urlcharsother \@ifnextchar [ {\mn@doi@}
  {\mn@doi@[]}}
\def\mn@doi@[#1]#2{\def\@tempa{#1}\ifx\@tempa\@empty \href
  {http://dx.doi.org/#2} {doi:#2}\else \href {http://dx.doi.org/#2} {#1}\fi
  \endgroup}
\def\mn@eprint#1#2{\mn@eprint@#1:#2::\@nil}
\def\mn@eprint@arXiv#1{\href {http://arxiv.org/abs/#1} {{\tt arXiv:#1}}}
\def\mn@eprint@dblp#1{\href {http://dblp.uni-trier.de/rec/bibtex/#1.xml}
  {dblp:#1}}
\def\mn@eprint@#1:#2:#3:#4\@nil{\def\@tempa {#1}\def\@tempb {#2}\def\@tempc
  {#3}\ifx \@tempc \@empty \let \@tempc \@tempb \let \@tempb \@tempa \fi \ifx
  \@tempb \@empty \def\@tempb {arXiv}\fi \@ifundefined
  {mn@eprint@\@tempb}{\@tempb:\@tempc}{\expandafter \expandafter \csname
  mn@eprint@\@tempb\endcsname \expandafter{\@tempc}}}

\bibitem[\protect\citeauthoryear{{Adamo} \& {Bastian}}{{Adamo} \&
  {Bastian}}{2018}]{Adamo_Bastian_18}
{Adamo} A.,  {Bastian} N.,  2018, {The Lifecycle of Clusters in Galaxies}.
p.~91, \mn@doi{10.1007/978-3-319-22801-3_4}

\bibitem[\protect\citeauthoryear{{Bastian}}{{Bastian}}{2008}]{Bastian_08}
{Bastian} N.,  2008, \mn@doi [\mnras] {10.1111/j.1365-2966.2008.13775.x}, \href
  {https://ui.adsabs.harvard.edu/abs/2008MNRAS.390..759B} {390, 759}

\bibitem[\protect\citeauthoryear{{Behroozi}, {Wechsler}  \&
  {Conroy}}{{Behroozi} et~al.}{2013}]{Behroozi_et_al_13}
{Behroozi} P.~S.,  {Wechsler} R.~H.,   {Conroy} C.,  2013, \mn@doi [\apj]
  {10.1088/0004-637X/770/1/57}, \href
  {https://ui.adsabs.harvard.edu/abs/2013ApJ...770...57B} {770, 57}

\bibitem[\protect\citeauthoryear{{Behroozi}, {Wechsler}, {Hearin}  \&
  {Conroy}}{{Behroozi} et~al.}{2019}]{Behroozi_et_al_19}
{Behroozi} P.,  {Wechsler} R.~H.,  {Hearin} A.~P.,   {Conroy} C.,  2019,
  \mn@doi [\mnras] {10.1093/mnras/stz1182}, \href
  {https://ui.adsabs.harvard.edu/abs/2019MNRAS.488.3143B} {488, 3143}

\bibitem[\protect\citeauthoryear{{Blakeslee}}{{Blakeslee}}{1999}]{blakeslee99}
{Blakeslee} J.~P.,  1999, \mn@doi [\aj] {10.1086/301052}, \href
  {https://ui.adsabs.harvard.edu/abs/1999AJ....118.1506B} {118, 1506}

\bibitem[\protect\citeauthoryear{{Blakeslee}, {Tonry}  \&
  {Metzger}}{{Blakeslee} et~al.}{1997}]{blakeslee97}
{Blakeslee} J.~P.,  {Tonry} J.~L.,   {Metzger} M.~R.,  1997, \mn@doi [\aj]
  {10.1086/118488}, \href
  {https://ui.adsabs.harvard.edu/abs/1997AJ....114..482B} {114, 482}

\bibitem[\protect\citeauthoryear{{Booth} \& {Schaye}}{{Booth} \&
  {Schaye}}{2009}]{Booth_and_Schaye_09}
{Booth} C.~M.,  {Schaye} J.,  2009, \mn@doi [\mnras]
  {10.1111/j.1365-2966.2009.15043.x}, \href
  {http://adsabs.harvard.edu/abs/2009MNRAS.398...53B} {398, 53}

\bibitem[\protect\citeauthoryear{{Boylan-Kolchin}}{{Boylan-Kolchin}}{2017}]{Boylan-Kolchin_17}
{Boylan-Kolchin} M.,  2017, \mn@doi [\mnras] {10.1093/mnras/stx2164}, \href
  {https://ui.adsabs.harvard.edu/abs/2017MNRAS.472.3120B} {472, 3120}

\bibitem[\protect\citeauthoryear{{Burkert} \& {Forbes}}{{Burkert} \&
  {Forbes}}{2020}]{burkert20}
{Burkert} A.,  {Forbes} D.~A.,  2020, \mn@doi [\aj] {10.3847/1538-3881/ab5b0e},
  \href {https://ui.adsabs.harvard.edu/abs/2020AJ....159...56B} {159, 56}

\bibitem[\protect\citeauthoryear{{Choksi} \& {Gnedin}}{{Choksi} \&
  {Gnedin}}{2019}]{Choksi_and_Gnedin_19}
{Choksi} N.,  {Gnedin} O.~Y.,  2019, \mn@doi [\mnras] {10.1093/mnras/stz2097},
  \href {https://ui.adsabs.harvard.edu/abs/2019MNRAS.488.5409C} {488, 5409}

\bibitem[\protect\citeauthoryear{{Cook} et~al.,}{{Cook}
  et~al.}{2012}]{Cook_et_al_12}
{Cook} D.~O.,  et~al., 2012, \mn@doi [\apj] {10.1088/0004-637X/751/2/100},
  \href {https://ui.adsabs.harvard.edu/abs/2012ApJ...751..100C} {751, 100}

\bibitem[\protect\citeauthoryear{{Crain} et~al.,}{{Crain}
  et~al.}{2015}]{Crain_et_al_15}
{Crain} R.~A.,  et~al., 2015, \mn@doi [\mnras] {10.1093/mnras/stv725}, \href
  {http://adsabs.harvard.edu/abs/2015MNRAS.450.1937C} {450, 1937}

\bibitem[\protect\citeauthoryear{{Davis}, {Efstathiou}, {Frenk}  \&
  {White}}{{Davis} et~al.}{1985}]{Davis_et_al_85}
{Davis} M.,  {Efstathiou} G.,  {Frenk} C.~S.,   {White} S.~D.~M.,  1985,
  \mn@doi [\apj] {10.1086/163168}, \href
  {http://ukads.nottingham.ac.uk/cgi-bin/nph-bib_query?bibcode=1985ApJ...292..371D&db_key=AST}
  {292, 371}

\bibitem[\protect\citeauthoryear{{Dolag}, {Borgani}, {Murante}  \&
  {Springel}}{{Dolag} et~al.}{2009}]{Dolag_et_al_09}
{Dolag} K.,  {Borgani} S.,  {Murante} G.,   {Springel} V.,  2009, \mn@doi
  [\mnras] {10.1111/j.1365-2966.2009.15034.x}, \href
  {http://adsabs.harvard.edu/abs/2009MNRAS.399..497D} {399, 497}

\bibitem[\protect\citeauthoryear{{El-Badry}, {Quataert}, {Weisz}, {Choksi}  \&
  {Boylan-Kolchin}}{{El-Badry} et~al.}{2019}]{El-Badry_et_al_19}
{El-Badry} K.,  {Quataert} E.,  {Weisz} D.~R.,  {Choksi} N.,   {Boylan-Kolchin}
  M.,  2019, \mn@doi [\mnras] {10.1093/mnras/sty3007}, \href
  {https://ui.adsabs.harvard.edu/abs/2019MNRAS.482.4528E} {482, 4528}

\bibitem[\protect\citeauthoryear{{Forbes}, {Alabi}, {Romanowsky}, {Brodie},
  {Strader}, {Usher}  \& {Pota}}{{Forbes} et~al.}{2016}]{forbes_et_al_16}
{Forbes} D.~A.,  {Alabi} A.,  {Romanowsky} A.~J.,  {Brodie} J.~P.,  {Strader}
  J.,  {Usher} C.,   {Pota} V.,  2016, \mn@doi [\mnras]
  {10.1093/mnrasl/slw015}, \href
  {https://ui.adsabs.harvard.edu/abs/2016MNRAS.458L..44F} {458, L44}

\bibitem[\protect\citeauthoryear{{Forbes} et~al.,}{{Forbes}
  et~al.}{2018a}]{forbes_review_18}
{Forbes} D.~A.,  et~al., 2018a, \mn@doi [Proceedings of the Royal Society of
  London Series A] {10.1098/rspa.2017.0616}, \href
  {https://ui.adsabs.harvard.edu/abs/2018RSPSA.47470616F} {474, 20170616}

\bibitem[\protect\citeauthoryear{{Forbes}, {Read}, {Gieles}  \&
  {Collins}}{{Forbes} et~al.}{2018b}]{Forbes_et_al_18}
{Forbes} D.~A.,  {Read} J.~I.,  {Gieles} M.,   {Collins} M. L.~M.,  2018b,
  \mn@doi [\mnras] {10.1093/mnras/sty2584}, \href
  {https://ui.adsabs.harvard.edu/abs/2018MNRAS.481.5592F} {481, 5592}

\bibitem[\protect\citeauthoryear{{Georgiev}, {Puzia}, {Goudfrooij}  \&
  {Hilker}}{{Georgiev} et~al.}{2010}]{Georgiev_et_al_10}
{Georgiev} I.~Y.,  {Puzia} T.~H.,  {Goudfrooij} P.,   {Hilker} M.,  2010,
  \mn@doi [\mnras] {10.1111/j.1365-2966.2010.16802.x}, \href
  {https://ui.adsabs.harvard.edu/abs/2010MNRAS.406.1967G} {406, 1967}

\bibitem[\protect\citeauthoryear{{Gieles} \& {Baumgardt}}{{Gieles} \&
  {Baumgardt}}{2008}]{Gieles_and_Baumgardt_08}
{Gieles} M.,  {Baumgardt} H.,  2008, \mn@doi [\mnras]
  {10.1111/j.1745-3933.2008.00515.x}, \href
  {https://ui.adsabs.harvard.edu/abs/2008MNRAS.389L..28G} {389, L28}

\bibitem[\protect\citeauthoryear{{Gnedin}, {Hernquist}  \& {Ostriker}}{{Gnedin}
  et~al.}{1999}]{Gnedin_Hernquist_and_Ostriker_99}
{Gnedin} O.~Y.,  {Hernquist} L.,   {Ostriker} J.~P.,  1999, \mn@doi [\apj]
  {10.1086/306910}, \href {http://adsabs.harvard.edu/abs/1999ApJ...514..109G}
  {514, 109}

\bibitem[\protect\citeauthoryear{{Harris} et~al.,}{{Harris}
  et~al.}{2014}]{Harris_et_al_14}
{Harris} W.~E.,  et~al., 2014, \mn@doi [\apj] {10.1088/0004-637X/797/2/128},
  \href {https://ui.adsabs.harvard.edu/abs/2014ApJ...797..128H} {797, 128}

\bibitem[\protect\citeauthoryear{{Harris}, {Blakeslee}  \& {Harris}}{{Harris}
  et~al.}{2017}]{Harris_et_al_17}
{Harris} W.~E.,  {Blakeslee} J.~P.,   {Harris} G. L.~H.,  2017, \mn@doi [\apj]
  {10.3847/1538-4357/836/1/67}, \href
  {https://ui.adsabs.harvard.edu/abs/2017ApJ...836...67H} {836, 67}

\bibitem[\protect\citeauthoryear{{Holtzman} et~al.,}{{Holtzman}
  et~al.}{1992}]{Holtzman_et_al_92}
{Holtzman} J.~A.,  et~al., 1992, \mn@doi [\aj] {10.1086/116094}, \href
  {https://ui.adsabs.harvard.edu/abs/1992AJ....103..691H} {103, 691}

\bibitem[\protect\citeauthoryear{{Hudson}, {Harris}  \& {Harris}}{{Hudson}
  et~al.}{2014}]{Hudson_et_al_14}
{Hudson} M.~J.,  {Harris} G.~L.,   {Harris} W.~E.,  2014, \mn@doi [\apjl]
  {10.1088/2041-8205/787/1/L5}, \href
  {https://ui.adsabs.harvard.edu/abs/2014ApJ...787L...5H} {787, L5}

\bibitem[\protect\citeauthoryear{{Hughes}, {Pfeffer}, {Martig}, {Bastian},
  {Crain}, {Kruijssen}  \& {Reina-Campos}}{{Hughes}
  et~al.}{2019}]{Hughes_et_al_19}
{Hughes} M.~E.,  {Pfeffer} J.,  {Martig} M.,  {Bastian} N.,  {Crain} R.~A.,
  {Kruijssen} J.~M.~D.,   {Reina-Campos} M.,  2019, \mn@doi [\mnras]
  {10.1093/mnras/sty2889}, \href
  {http://adsabs.harvard.edu/abs/2019MNRAS.482.2795H} {482, 2795}

\bibitem[\protect\citeauthoryear{{Hughes}, {Pfeffer}, {Martig}, {Reina-Campos},
  {Bastian}, {Crain}  \& {Kruijssen}}{{Hughes} et~al.}{2020}]{Hughes_et_al_20}
{Hughes} M.~E.,  {Pfeffer} J.~L.,  {Martig} M.,  {Reina-Campos} M.,  {Bastian}
  N.,  {Crain} R.~A.,   {Kruijssen} J.~M.~D.,  2020, \mn@doi [\mnras]
  {10.1093/mnras/stz3341}, \href
  {https://ui.adsabs.harvard.edu/abs/2020MNRAS.491.4012H} {491, 4012}

\bibitem[\protect\citeauthoryear{{Jahnke} \& {Macci{\`o}}}{{Jahnke} \&
  {Macci{\`o}}}{2011}]{jahnke_and_maccio_11}
{Jahnke} K.,  {Macci{\`o}} A.~V.,  2011, \mn@doi [\apj]
  {10.1088/0004-637X/734/2/92}, \href
  {https://ui.adsabs.harvard.edu/abs/2011ApJ...734...92J} {734, 92}

\bibitem[\protect\citeauthoryear{{Johnson} et~al.,}{{Johnson}
  et~al.}{2017}]{Johnson_et_al_17}
{Johnson} T.~L.,  et~al., 2017, \mn@doi [\apjl] {10.3847/2041-8213/aa7516},
  \href {https://ui.adsabs.harvard.edu/abs/2017ApJ...843L..21J} {843, L21}

\bibitem[\protect\citeauthoryear{{Jord{\'a}n} et~al.,}{{Jord{\'a}n}
  et~al.}{2007}]{Jordan_et_al_07}
{Jord{\'a}n} A.,  et~al., 2007, \mn@doi [\apjs] {10.1086/516840}, \href
  {http://adsabs.harvard.edu/abs/2007ApJS..171..101J} {171, 101}

\bibitem[\protect\citeauthoryear{{Kruijssen}}{{Kruijssen}}{2012}]{Kruijssen_12}
{Kruijssen} J.~M.~D.,  2012, \mn@doi [\mnras]
  {10.1111/j.1365-2966.2012.21923.x}, \href
  {http://adsabs.harvard.edu/abs/2012MNRAS.426.3008K} {426, 3008}

\bibitem[\protect\citeauthoryear{{Kruijssen}}{{Kruijssen}}{2014}]{Kruijssen_14}
{Kruijssen} J.~M.~D.,  2014, \mn@doi [Classical and Quantum Gravity]
  {10.1088/0264-9381/31/24/244006}, \href
  {http://adsabs.harvard.edu/abs/2014CQGra..31x4006K} {31, 244006}

\bibitem[\protect\citeauthoryear{{Kruijssen}}{{Kruijssen}}{2015}]{Kruijssen_15}
{Kruijssen} J.~M.~D.,  2015, \mn@doi [\mnras] {10.1093/mnras/stv2026}, \href
  {http://adsabs.harvard.edu/abs/2015MNRAS.454.1658K} {454, 1658}

\bibitem[\protect\citeauthoryear{{Kruijssen}, {Pelupessy}, {Lamers}, {Portegies
  Zwart}  \& {Icke}}{{Kruijssen} et~al.}{2011}]{Kruijssen_et_al_11}
{Kruijssen} J.~M.~D.,  {Pelupessy} F.~I.,  {Lamers} H.~J.~G.~L.~M.,  {Portegies
  Zwart} S.~F.,   {Icke} V.,  2011, \mn@doi [\mnras]
  {10.1111/j.1365-2966.2011.18467.x}, \href
  {http://adsabs.harvard.edu/abs/2011MNRAS.414.1339K} {414, 1339}

\bibitem[\protect\citeauthoryear{{Kruijssen}, {Pfeffer}, {Crain}  \&
  {Bastian}}{{Kruijssen} et~al.}{2019a}]{Kruijssen_et_al_19a}
{Kruijssen} J.~M.~D.,  {Pfeffer} J.~L.,  {Crain} R.~A.,   {Bastian} N.,  2019a,
  \mn@doi [\mnras] {10.1093/mnras/stz968}, \href
  {https://ui.adsabs.harvard.edu/abs/2019MNRAS.486.3134K} {486, 3134}

\bibitem[\protect\citeauthoryear{{Kruijssen}, {Pfeffer}, {Reina-Campos},
  {Crain}  \& {Bastian}}{{Kruijssen} et~al.}{2019b}]{Kruijssen_et_al_19b}
{Kruijssen} J.~M.~D.,  {Pfeffer} J.~L.,  {Reina-Campos} M.,  {Crain} R.~A.,
  {Bastian} N.,  2019b, \mn@doi [\mnras] {10.1093/mnras/sty1609}, \href
  {https://ui.adsabs.harvard.edu/abs/2019MNRAS.486.3180K} {486, 3180}

\bibitem[\protect\citeauthoryear{{Kruijssen} et~al.,}{{Kruijssen}
  et~al.}{2020}]{Kruijssen_et_al_20}
{Kruijssen} J.~M.~D.,  et~al., 2020, \mnras~submitted, arXiv:2003.01119

\bibitem[\protect\citeauthoryear{{Mackey} et~al.,}{{Mackey}
  et~al.}{2019}]{Mackey_et_al_19}
{Mackey} D.,  et~al., 2019, \mn@doi [\nat] {10.1038/s41586-019-1597-1}, \href
  {https://ui.adsabs.harvard.edu/abs/2019Natur.574...69M} {574, 69}

\bibitem[\protect\citeauthoryear{{Mistani} et~al.,}{{Mistani}
  et~al.}{2016}]{Mistani_et_al_16}
{Mistani} P.~A.,  et~al., 2016, \mn@doi [\mnras] {10.1093/mnras/stv2435}, \href
  {https://ui.adsabs.harvard.edu/abs/2016MNRAS.455.2323M} {455, 2323}

\bibitem[\protect\citeauthoryear{{Moster}, {Naab}  \& {White}}{{Moster}
  et~al.}{2013}]{Moster_et_al_13}
{Moster} B.~P.,  {Naab} T.,   {White} S.~D.~M.,  2013, \mn@doi [\mnras]
  {10.1093/mnras/sts261}, \href
  {http://adsabs.harvard.edu/abs/2013MNRAS.428.3121M} {428, 3121}

\bibitem[\protect\citeauthoryear{{Moster}, {Naab}  \& {White}}{{Moster}
  et~al.}{2018}]{moster_et_al_18}
{Moster} B.~P.,  {Naab} T.,   {White} S. D.~M.,  2018, \mn@doi [\mnras]
  {10.1093/mnras/sty655}, \href
  {https://ui.adsabs.harvard.edu/abs/2018MNRAS.477.1822M} {477, 1822}

\bibitem[\protect\citeauthoryear{{Peng} et~al.,}{{Peng}
  et~al.}{2008}]{Peng_et_al_08}
{Peng} E.~W.,  et~al., 2008, \mn@doi [\apj] {10.1086/587951}, \href
  {http://adsabs.harvard.edu/abs/2008ApJ...681..197P} {681, 197}

\bibitem[\protect\citeauthoryear{{Pfeffer}, {Kruijssen}, {Crain}  \&
  {Bastian}}{{Pfeffer} et~al.}{2018}]{Pfeffer_et_al_18}
{Pfeffer} J.,  {Kruijssen} J.~M.~D.,  {Crain} R.~A.,   {Bastian} N.,  2018,
  \mn@doi [\mnras] {10.1093/mnras/stx3124}, \href
  {http://adsabs.harvard.edu/abs/2018MNRAS.475.4309P} {475, 4309}

\bibitem[\protect\citeauthoryear{{Pfeffer}, {Bastian}, {Kruijssen},
  {Reina-Campos}, {Crain}  \& {Usher}}{{Pfeffer}
  et~al.}{2019}]{Pfeffer_et_al_19b}
{Pfeffer} J.,  {Bastian} N.,  {Kruijssen} J.~M.~D.,  {Reina-Campos} M.,
  {Crain} R.~A.,   {Usher} C.,  2019, \mn@doi [\mnras] {10.1093/mnras/stz2721},
  \href {https://ui.adsabs.harvard.edu/abs/2019MNRAS.490.1714P} {490, 1714}

\bibitem[\protect\citeauthoryear{{Planck Collaboration} et~al.,}{{Planck
  Collaboration} et~al.}{2014}]{Planck_2014_paperXVI}
{Planck Collaboration} et~al., 2014, \mn@doi [\aap]
  {10.1051/0004-6361/201321591}, \href
  {http://adsabs.harvard.edu/abs/2014A%26A...571A..16P} {571, A16}

\bibitem[\protect\citeauthoryear{{Prieto} \& {Gnedin}}{{Prieto} \&
  {Gnedin}}{2008}]{Prieto_and_Gnedin_08}
{Prieto} J.~L.,  {Gnedin} O.~Y.,  2008, \mn@doi [\apj] {10.1086/591777}, \href
  {http://adsabs.harvard.edu/abs/2008ApJ...689..919P} {689, 919}

\bibitem[\protect\citeauthoryear{{Reina-Campos} \& {Kruijssen}}{{Reina-Campos}
  \& {Kruijssen}}{2017}]{Reina-Campos_and_Kruijssen_17}
{Reina-Campos} M.,  {Kruijssen} J.~M.~D.,  2017, \mn@doi [\mnras]
  {10.1093/mnras/stx790}, \href
  {http://adsabs.harvard.edu/abs/2017MNRAS.469.1282R} {469, 1282}

\bibitem[\protect\citeauthoryear{{Reina-Campos}, {Kruijssen}, {Pfeffer},
  {Bastian}  \& {Crain}}{{Reina-Campos} et~al.}{2018}]{Reina-Campos_et_al_18}
{Reina-Campos} M.,  {Kruijssen} J.~M.~D.,  {Pfeffer} J.,  {Bastian} N.,
  {Crain} R.~A.,  2018, \mn@doi [\mnras] {10.1093/mnras/sty2451}, \href
  {http://adsabs.harvard.edu/abs/2018MNRAS.481.2851R} {481, 2851}

\bibitem[\protect\citeauthoryear{{Reina-Campos}, {Kruijssen}, {Pfeffer},
  {Bastian}  \& {Crain}}{{Reina-Campos} et~al.}{2019}]{Reina-Campos_et_al_19}
{Reina-Campos} M.,  {Kruijssen} J.~M.~D.,  {Pfeffer} J.~L.,  {Bastian} N.,
  {Crain} R.~A.,  2019, \mn@doi [\mnras] {10.1093/mnras/stz1236}, \href
  {https://ui.adsabs.harvard.edu/abs/2019MNRAS.486.5838R} {486, 5838}

\bibitem[\protect\citeauthoryear{{Reina-Campos}, {Hughes}, {Kruijssen},
  {Pfeffer}, {Bastian}, {Crain}, {Koch}  \& {Grebel}}{{Reina-Campos}
  et~al.}{2020}]{Reina-Campos_et_al_20}
{Reina-Campos} M.,  {Hughes} M.~E.,  {Kruijssen} J.~M.~D.,  {Pfeffer} J.~L.,
  {Bastian} N.,  {Crain} R.~A.,  {Koch} A.,   {Grebel} E.~K.,  2020, \mn@doi
  [\mnras] {10.1093/mnras/staa483}, \href
  {https://ui.adsabs.harvard.edu/abs/2020MNRAS.493.3422R} {493, 3422}

\bibitem[\protect\citeauthoryear{{Rosas-Guevara} et~al.,}{{Rosas-Guevara}
  et~al.}{2015}]{Rosas-Guevara_et_al_15}
{Rosas-Guevara} Y.~M.,  et~al., 2015, \mn@doi [\mnras] {10.1093/mnras/stv2056},
  \href {http://adsabs.harvard.edu/abs/2015MNRAS.454.1038R} {454, 1038}

\bibitem[\protect\citeauthoryear{{Schaye} \& {Dalla Vecchia}}{{Schaye} \&
  {Dalla Vecchia}}{2008}]{Schaye_and_Dalla_Vecchia_08}
{Schaye} J.,  {Dalla Vecchia} C.,  2008, \mn@doi [\mnras]
  {10.1111/j.1365-2966.2007.12639.x}, \href
  {http://adsabs.harvard.edu/abs/2008MNRAS.383.1210S} {383, 1210}

\bibitem[\protect\citeauthoryear{{Schaye} et~al.,}{{Schaye}
  et~al.}{2015}]{Schaye_et_al_15}
{Schaye} J.,  et~al., 2015, \mn@doi [\mnras] {10.1093/mnras/stu2058}, \href
  {http://adsabs.harvard.edu/abs/2015MNRAS.446..521S} {446, 521}

\bibitem[\protect\citeauthoryear{{Schechter}}{{Schechter}}{1976}]{Schechter_76}
{Schechter} P.,  1976, \mn@doi [\apj] {10.1086/154079}, \href
  {http://adsabs.harvard.edu/abs/1976ApJ...203..297S} {203, 297}

\bibitem[\protect\citeauthoryear{{Schweizer} \& {Seitzer}}{{Schweizer} \&
  {Seitzer}}{1998}]{Schweizer_and_Seitzer_98}
{Schweizer} F.,  {Seitzer} P.,  1998, \mn@doi [\aj] {10.1086/300616}, \href
  {https://ui.adsabs.harvard.edu/abs/1998AJ....116.2206S} {116, 2206}

\bibitem[\protect\citeauthoryear{{Spitler} \& {Forbes}}{{Spitler} \&
  {Forbes}}{2009}]{Spitler_and_Forbes_09}
{Spitler} L.~R.,  {Forbes} D.~A.,  2009, \mn@doi [\mnras]
  {10.1111/j.1745-3933.2008.00567.x}, \href
  {https://ui.adsabs.harvard.edu/abs/2009MNRAS.392L...1S} {392, L1}

\bibitem[\protect\citeauthoryear{{Springel}}{{Springel}}{2005}]{Springel_05}
{Springel} V.,  2005, \mn@doi [\mnras] {10.1111/j.1365-2966.2005.09655.x},
  \href
  {http://adsabs.harvard.edu/cgi-bin/nph-bib_query?bibcode=2005MNRAS.364.1105S&db_key=AST}
  {364, 1105}

\bibitem[\protect\citeauthoryear{{Springel}, {White}, {Tormen}  \&
  {Kauffmann}}{{Springel} et~al.}{2001}]{Springel_et_al_01}
{Springel} V.,  {White} S.~D.~M.,  {Tormen} G.,   {Kauffmann} G.,  2001,
  \mn@doi [\mnras] {10.1046/j.1365-8711.2001.04912.x}, \href
  {http://adsabs.harvard.edu/abs/2001MNRAS.328..726S} {328, 726}

\bibitem[\protect\citeauthoryear{{Usher}, {Pfeffer}, {Bastian}, {Kruijssen},
  {Crain}  \& {Reina-Campos}}{{Usher} et~al.}{2018}]{Usher_et_al_18}
{Usher} C.,  {Pfeffer} J.,  {Bastian} N.,  {Kruijssen} J.~M.~D.,  {Crain}
  R.~A.,   {Reina-Campos} M.,  2018, \mn@doi [\mnras] {10.1093/mnras/sty1895},
  \href {http://adsabs.harvard.edu/abs/2018MNRAS.480.3279U} {480, 3279}

\bibitem[\protect\citeauthoryear{{Usher}, {Brodie}, {Forbes}, {Romanowsky},
  {Strader}, {Pfeffer}  \& {Bastian}}{{Usher} et~al.}{2019}]{Usher_et_al_19}
{Usher} C.,  {Brodie} J.~P.,  {Forbes} D.~A.,  {Romanowsky} A.~J.,  {Strader}
  J.,  {Pfeffer} J.,   {Bastian} N.,  2019, \mn@doi [\mnras]
  {10.1093/mnras/stz2596}, \href
  {https://ui.adsabs.harvard.edu/abs/2019MNRAS.490..491U} {490, 491}

\bibitem[\protect\citeauthoryear{{Vanzella} et~al.,}{{Vanzella}
  et~al.}{2017}]{Vanzella_et_al_17}
{Vanzella} E.,  et~al., 2017, \mn@doi [\mnras] {10.1093/mnras/stx351}, \href
  {https://ui.adsabs.harvard.edu/abs/2017MNRAS.467.4304V} {467, 4304}

\bibitem[\protect\citeauthoryear{{Wiersma}, {Schaye}  \& {Smith}}{{Wiersma}
  et~al.}{2009a}]{Wiersma_Schaye_and_Smith_09}
{Wiersma} R.~P.~C.,  {Schaye} J.,   {Smith} B.~D.,  2009a, \mn@doi [\mnras]
  {10.1111/j.1365-2966.2008.14191.x}, \href
  {http://adsabs.harvard.edu/abs/2009MNRAS.393...99W} {393, 99}

\bibitem[\protect\citeauthoryear{{Wiersma}, {Schaye}, {Theuns}, {Dalla Vecchia}
   \& {Tornatore}}{{Wiersma} et~al.}{2009b}]{Wiersma_et_al_09}
{Wiersma} R.~P.~C.,  {Schaye} J.,  {Theuns} T.,  {Dalla Vecchia} C.,
  {Tornatore} L.,  2009b, \mn@doi [\mnras] {10.1111/j.1365-2966.2009.15331.x},
  \href {http://adsabs.harvard.edu/abs/2009MNRAS.399..574W} {399, 574}

\bibitem[\protect\citeauthoryear{{van de Voort}, {Schaye}, {Booth}  \& {Dalla
  Vecchia}}{{van de Voort} et~al.}{2011}]{van_de_voort_11}
{van de Voort} F.,  {Schaye} J.,  {Booth} C.~M.,   {Dalla Vecchia} C.,  2011,
  \mn@doi [\mnras] {10.1111/j.1365-2966.2011.18896.x}, \href
  {https://ui.adsabs.harvard.edu/abs/2011MNRAS.415.2782V} {415, 2782}

\makeatother
\end{thebibliography}

%
%
%
%
%
%
%
%
%
%
%
%



\appendix
\section{GC Selection}
\label{sec:gc_selection}

In this study, we determine the total mass of the GC population (\mgc) of a galaxy as the sum of the top two decades of mass within the cluster system.  We adopt this approach rather than using a fixed mass limit (e.g., all mass above $10^5$~\msun), because it is more likely to approximate GC selection in observational surveys.  Observations of GC systems have a wide range of completeness limits, with some Local Group studies being complete to below $10^4$~\msun\ \citep[e.g.][and references therein]{Forbes_et_al_18}, in fact in some cases the most massive cluster is below $10^4$~\msun.  In these and similar cases, using GCs only above a fixed limit will remove a large fraction of the GC mass within these systems.   Other studies, focusing on massive central galaxies at $100-200$~Mpc, are complete only to around the turnover of the GC mass or luminosity function or above \citep[$10^5$-$10^6$~\msun,][]{Harris_et_al_14}.  One exception to our definition above is that we do not let the lower mass limit of the integration go above $2\times 10^5$~\msun, similar to the GC turnover mass in $>10^{11}$~\msun\ galaxies \citep{Jordan_et_al_07,Harris_et_al_14}. 

We do not make any cuts on metallicity, age or galactocentric distance unless explicitly mentioned in the text.

\begin{figure}
\centering
\includegraphics[width=8cm]{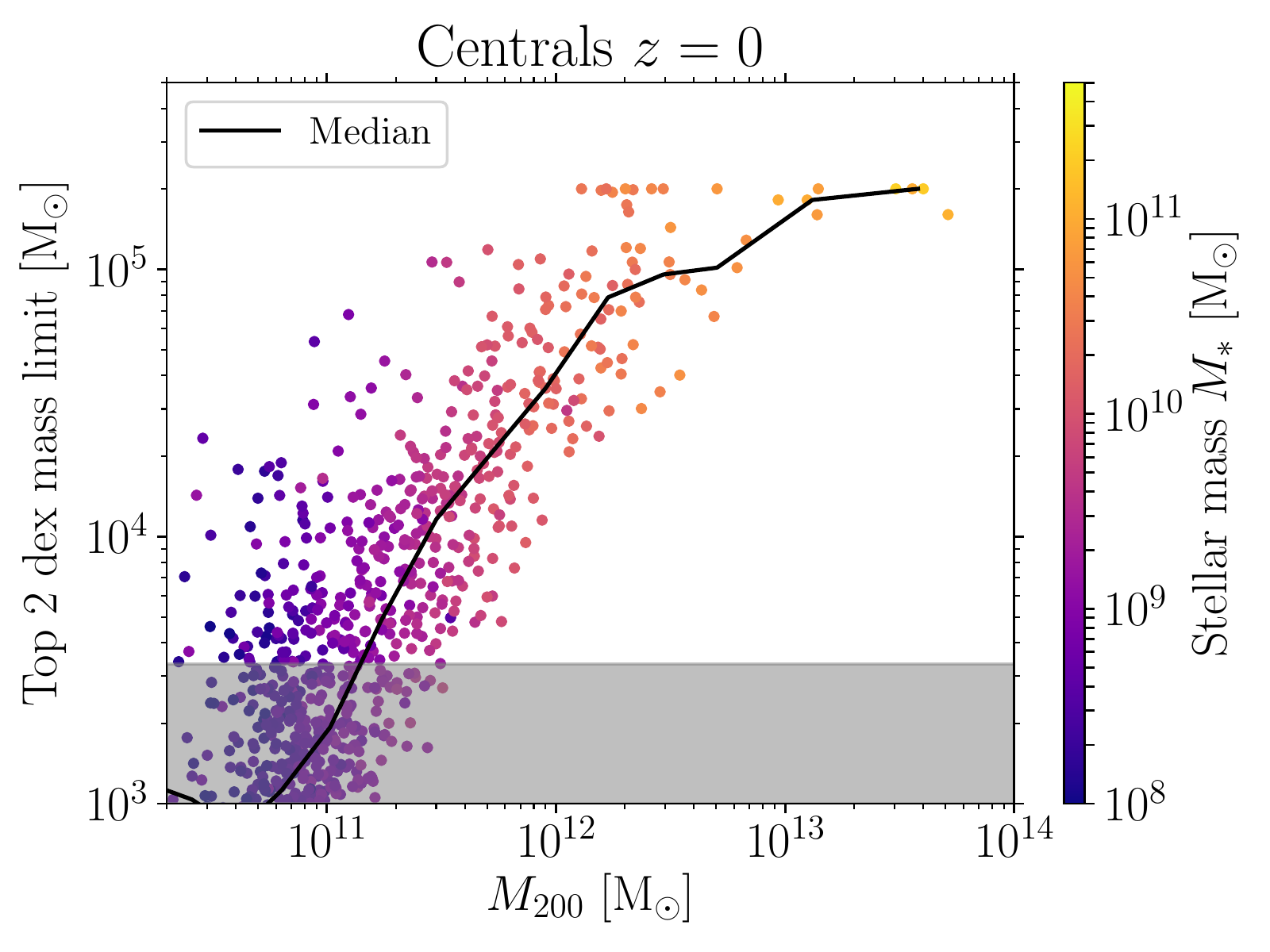}
\caption{The lower limit of the mass range of our integration for each of the cluster populations in our study.  For the total mass in GCs (\mgc), we restrict our integration to only the top two decades in mass of the cluster population (or slightly more if the lower limit of the integration exceeds $2\times10^5$~\msun).  The solid black line shows the median mass limit as a function of $M_{200}$.  The shaded region shows where the lower mass limit goes below our minimum initial cluster mass limit of $5\times10^3$~\msun\ (descreased by a factor $0.67$ to approximately account for stellar-evolutionary mass loss).} 
\label{fig:top_2dex}
\end{figure}

In Fig.~\ref{fig:top_2dex} we show the resulting lower mass limit for \mgc\ for each of the galaxies in our sample.
The median mass limit is a strong function of the galaxy's stellar or halo mass, reaching $<2\times 10^3$~\msun\ for haloes with $M_{200}<10^{11}$~\msun\ and $>10^5$~\msun\ for haloes $M_{200}>2\times 10^{12}$~\msun. At $M_{200}> 10^{13}$~\msun\ the mass limits reach our adopted minimum of $2\times 10^5$~\msun.

\begin{figure}
\centering
\includegraphics[width=8cm]{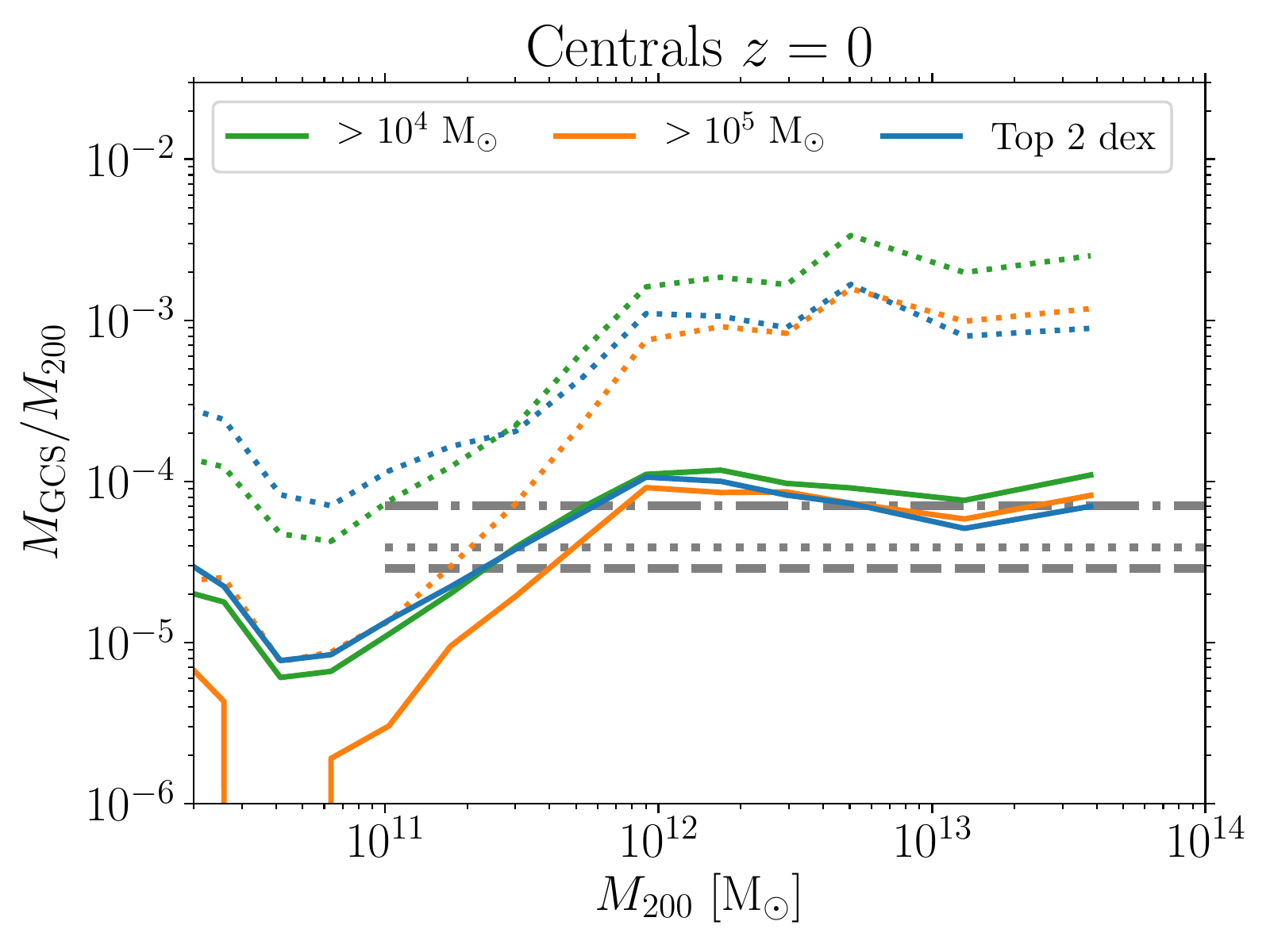}
\caption{The \mgc-\mhalo\ relation for different definitions of \mgc\ using different lower GC mass limits.  The dotted lines show the results for the initial GC masses (with a lower limit $1.5$ times that of the limit for evolved cluster masses, to account for stellar-evolutionary mass loss) and the solid lines show the results for $z=0$ and including cluster disruption.  Due to the disruption of low-mass clusters the results are not strongly affected by the choice in the limit of defining GC mass within the simulations.} 
\label{fig:gc_def}
\end{figure}

We explore the results of adopting a fixed mass limit, which are qualitatively largely similar to our variable limit, in Fig.~\ref{fig:gc_def}.  
At halo masses $M_{200}>10^{12}$~\msun, the our results are insensitive to the mass limit adopted.
At halo masses $M_{200}<10^{12}$~\msun, adopting a mass limit of $10^5$~\msun\ would impart an artificial downturn in the \mgc-\mhalo\ relation, due to low mass galaxies forming few very massive GCs.
Note that the upturn at $M_{200} \lesssim 4\times10^{10}$~\msun\ is due to the fixed stellar mass selection of $M_\ast>10^8$~\msun.

\bsp
\label{lastpage}
\end{document}